\documentclass[twocolumn]{aastex631}

\defcitealias{GC21}{GC21}

\shorttitle{The LMC's effect on MW satellite orbital poles}
\shortauthors{Pawlowski et al.}
\graphicspath{{./}{figures_v2/}}

\begin{document}

\title{On the Effect of the Large Magellanic Cloud on the Orbital Poles of Milky Way Satellite Galaxies}

\correspondingauthor{Marcel S. Pawlowski}
\email{mpawlowski@aip.de}

\author[0000-0002-9197-9300]{Marcel S. Pawlowski}
\affiliation{Leibniz-Institut f\"ur Astrophysik Potsdam (AIP), An der Sternwarte 16, D-14482 Potsdam, Germany}

\author{Pierre-Antoine Oria}
\affiliation{Universit\'{e} de Strasbourg, CNRS, Observatoire astronomique de Strasbourg, UMR 7550, F-67000 Strasbourg, France}

\author[0000-0001-6469-8805]{Salvatore Taibi}
\affiliation{Leibniz-Institut f\"ur Astrophysik Potsdam (AIP), An der Sternwarte 16, D-14482 Potsdam, Germany}

\author[0000-0003-3180-9825]{Benoit Famaey}
\affiliation{Universit\'{e} de Strasbourg, CNRS, Observatoire astronomique de Strasbourg, UMR 7550, F-67000 Strasbourg, France}

\author[0000-0002-3292-9709]{Rodrigo Ibata}
\affiliation{Universit\'{e} de Strasbourg, CNRS, Observatoire astronomique de Strasbourg, UMR 7550, F-67000 Strasbourg, France}



\begin{abstract}
The reflex motion and distortion of the Milky Way (MW) halo caused by the infall of a massive Large Magellanic Cloud (LMC) has been demonstrated to result in an excess of orbital poles of dark matter halo particles towards the LMC orbital pole. This was suggested to help explain the observed preference of MW satellite galaxies to co-orbit along the Vast Polar Structure (VPOS). We test this idea by correcting the positions and velocities of the MW satellites for the Galactocentric-distance-dependent shifts inferred from a LMC-infall simulation. While this should substantially reduce the observed clustering of orbital poles if it were mainly caused by the LMC, we instead find that the strong clustering remains preserved. We confirm the initial study's main result with our simulation of an MW-LMC-like interaction, and use it to identify two reasons why this scenario is unable to explain the VPOS: (1) the orbital pole density enhancement in our simulation is very mild ($\sim10\%$\ within 50-250\,kpc) compared to the observed enhancement ($\sim$220-300\%), and (2) it is very sensitive to the specific angular momenta (AM) of the simulation particles, with higher AM particles being affected the least. Particles in simulated dark matter halos tend to follow more radial orbits (lower AM), so their orbital poles are more easily affected by small offsets in position and velocity caused by an LMC infall than objects with more tangential velocity (higher AM), such as the observed dwarf galaxies surrounding the MW. The origin of the VPOS thus remains unexplained.
\end{abstract}



\section{Introduction} \label{sec:intro}

\begin{figure*}
\gridline{
\fig{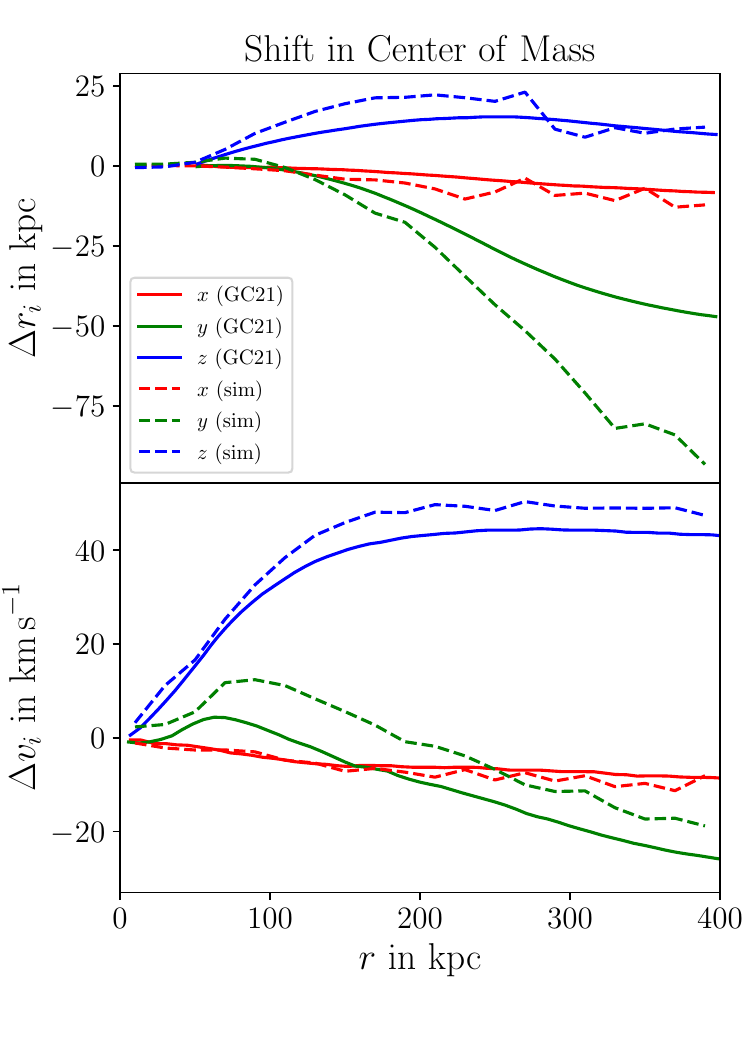}{0.375\textwidth}{(a)}
\fig{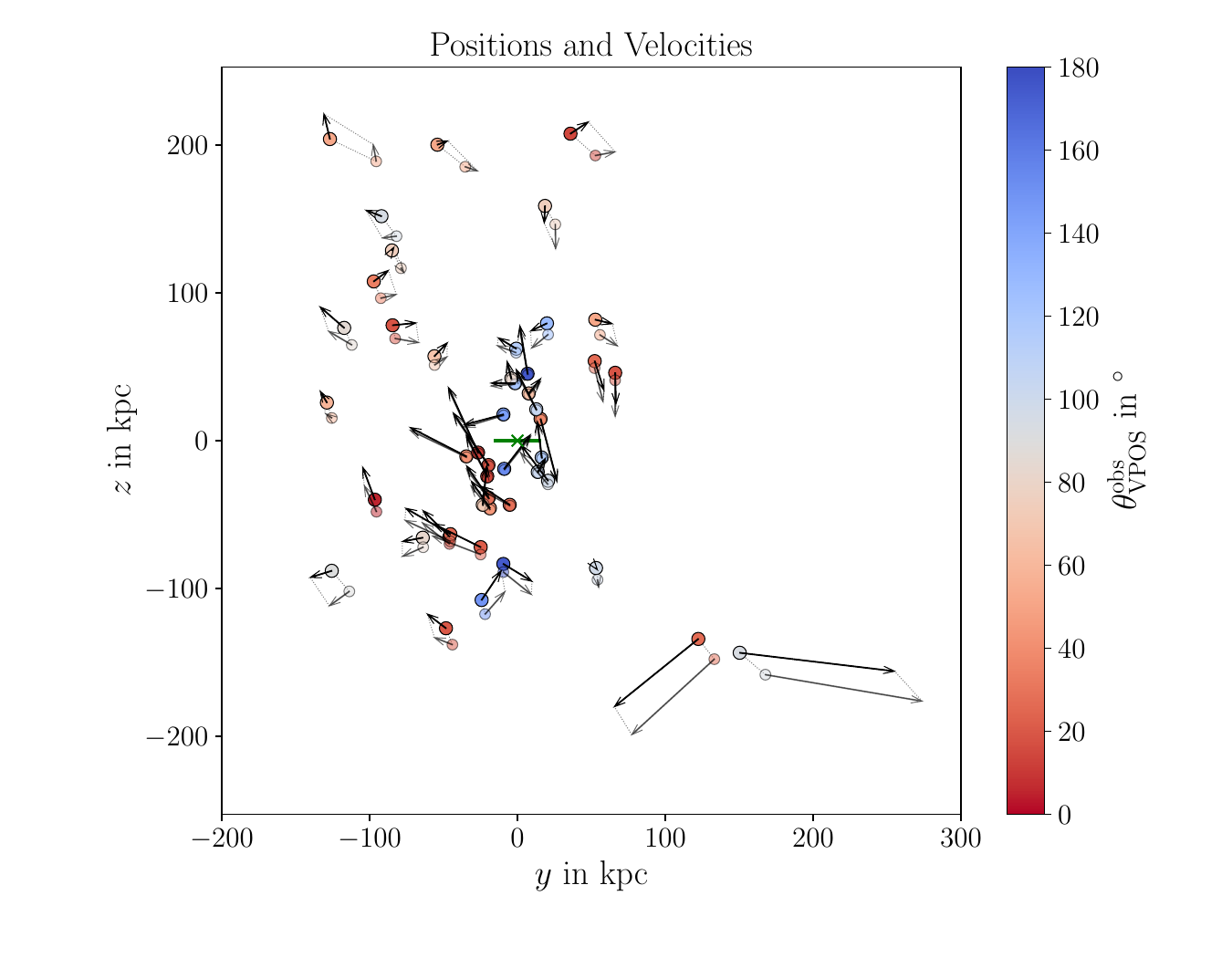}{0.675\textwidth}{(b)}
}
\caption{
Effect of an LMC-induced shift in the COM positions and velocities on the observed MW dwarfs. 
Panel (a) shows the shifts in the center-of-mass position $\Delta r_i$\ and velocity $\Delta v_i$\ for simulation particles in spherical shells of distance $r$\ around their host galaxy center, in three Cartesian coordinates $i \in \{x, y, z\}$, with $x$ measured along the Galactic Center-Sun direction, $y$ in the direction of Galactic rotation at the Sun, and $z$ towards the North Galactic Pole. The results reported by \citetalias{GC21} are shown (solid lines) and compared to our final simulation snapshot (dashed lines). Panel (b) shows the positions and velocities (as vectors) of the MW satellites. They are color-coded by the observed alignment angle $\theta_\mathrm{VPOS}^\mathrm{obs}$\ of their orbital pole directions with the VPOS normal (red is co-orbiting). Smaller, fainter symbols indicate the position after applying the \citetalias{GC21} shifts from panel (a) in reverse. The $y-z$\ plane is displayed, in which the majority of the shift takes place and which aligns well with the VPOS. The position and orientation of the edge-on MW is indicated in green.
\label{fig:shifts}}
\end{figure*}

The phase-space correlation of satellite galaxies around their hosts is currently one of the most pressing challenges for our understanding of galaxy formation \citep{2017ARA&A..55..343B, 2018MPLA...3330004P}. In short, a large fraction of dwarf satellite galaxies of the Milky Way (MW) and M31 are distributed within flattened and kinematically coherent structures that are at odds with the expected phase-space distribution of their associated dark matter sub-halos \citep{2005A&A...431..517K, 2013Natur.493...62I}. In the MW, this structure is called the Vast Polar Structure (VPOS, \citealt{2012MNRAS.423.1109P}): it is perpendicular to the galactic disc, has a root-mean-square thickness of only $\sim$25 kpc for a 10 times larger spatial extension, and its normal vector is aligned with the orbital poles of 50\% to 75\% of the galaxies spatially located in it \citep{2021ApJ...916....8L}.

Similar planar structures have been found outside of the Local Group \citep[e.g.][]{2018Sci...359..534M, 2021A&A...645L...5M, 2021ApJ...917L..18P, 2021A&A...652A..48M, 2021arXiv210810189H}, and are calling for an explanation. Many theoretical scenarios for the formation of these satellite-plane structures have been proposed over the years \citep[for a review, see][]{2018MPLA...3330004P}, but thus far all have failed to clearly reproduce the extreme phase-space distributions observed around real galaxies. Recently, a new possible explanation for the VPOS around the MW has been put forward by \citet{GC21} (hereafter \citetalias{GC21}). It partly relies on the old observation by \citet{1976MNRAS.174..695L} and \citet{1976RGOB..182..241K} that the Large Magellanic Cloud (LMC) and the Magellanic Stream, align with and orbit along the VPOS.

In recent years, an array of evidence has accumulated for a more massive LMC than initially thought \citep{2013ApJ...764..161K, 2016MNRAS.456L..54P, 2019A&A...623A.129F, 2020MNRAS.495.2554E, 2021NatAs...5..251P}, with a total mass at infall of at least $1.5 \times 10^{11} \, {\rm M}_\odot$. Such a massive LMC should necessarily cause a shift in the orbital barycenter of the MW–LMC system compared to the center of the MW's disk, which could influence the kinematics of satellite galaxies.

\citetalias{GC21} demonstrated with a numerical simulation of a massive LMC-analog interacting with a MW-like host that the resulting reflex motion and offset in overall center-of-mass conspire to induce an overdensity of orbital poles in the dark matter particles that constitute the dark halo of the MW model. The direction of this overdensity aligns with the orbital pole of their LMC analog. As the orbital pole of the observed LMC is well aligned with the observed VPOS and its associated overdensity of satellite galaxy orbital poles, \citetalias{GC21} suggest that this LMC-induced effect might help explain the presence of the VPOS.

We here go beyond the original study and test whether this proposed mechanism can have a sufficient effect on the inferred orbital poles of the {\it observed} MW satellite system.

\section{Effect on Milky Way Satellites} \label{sec:MWsats}

\begin{figure}
\gridline{
\fig{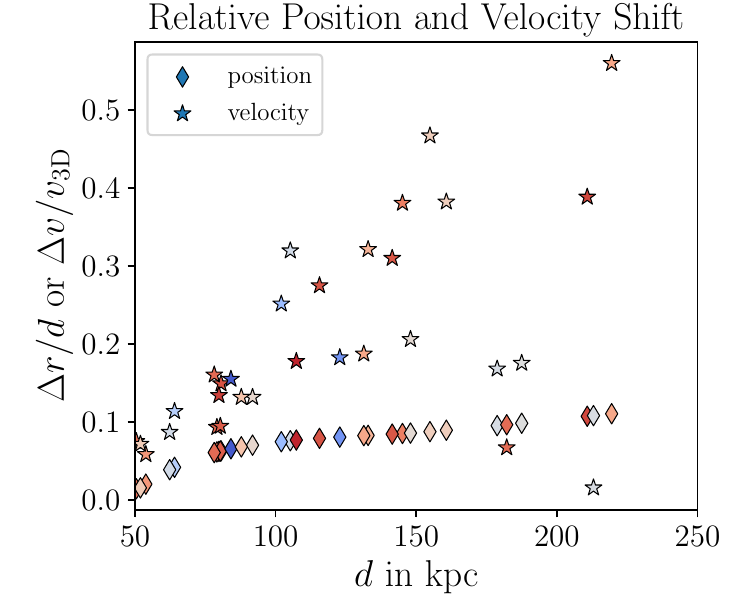}{0.4\textwidth}{(a)}
}
\gridline{
\fig{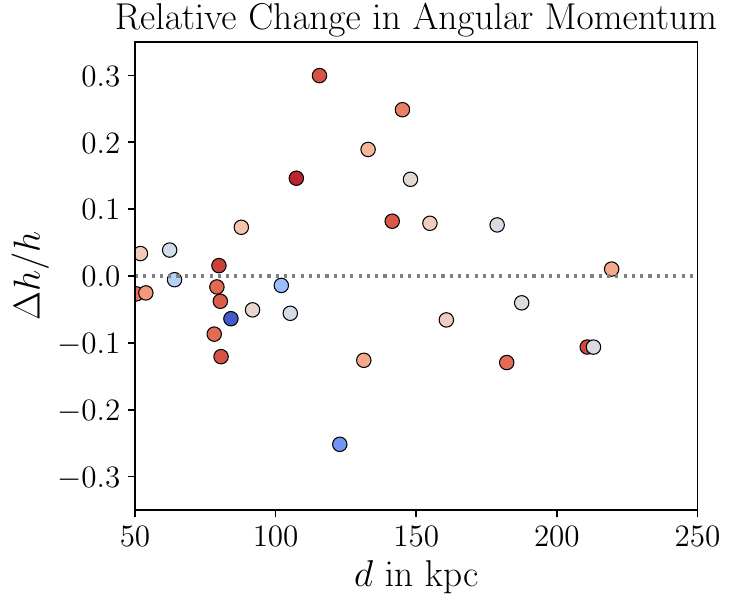}{0.4\textwidth}{(b)}
}
\gridline{
\fig{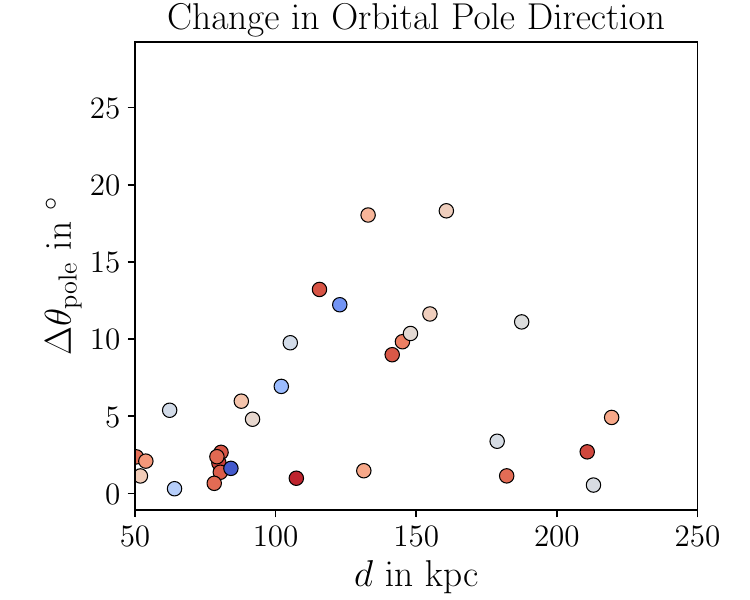}{0.4\textwidth}{(c)}
}
\caption{
Panels (a) to (c) show the relative effect that the COM shifts have (see Fig. \ref{fig:shifts}) on the position and velocity, angular momentum, and direction of orbital pole, respectively. Here and throughout the following figures the MW dwarfs are color-coded as in panel (b) of Fig.~\ref{fig:shifts}.
\label{fig:shifts2}}
\end{figure}

\citetalias{GC21} argue that the orbital pole enhancement in their simulation can be reproduced by shifting the positions and velocities of simulation particles in the initial conditions. Effectively, this means that for different radial shells, different reference centers-of-mass (COM) are adopted, mimicking the perturbation to the MW's dark matter halo by the infall of a massive LMC. They have demonstrated that this approach accounts for both the degree and direction of the orbital pole overdensity of dark matter particles in their simulation, making a convincing argument for the validity of their model.

We adopt this model but invert the approach to estimate to what degree the infall of the LMC can be responsible for the observed clustering of satellite orbital poles. If adding shifts to an isotropic initial distribution reproduces the LMC's influence, then subtracting those shifts from the observed MW dwarf galaxies, which are exposed to the LMC's influence, should substantially reduce the observed orbital pole overdensity if its main cause is the gravitational influence of the LMC. We note that this approach does only take the reflex motion and COM offset into account, but not any orbital pole enhancement due to the wake of the LMC or its direct torque effects on MW satellites galaxies. As recently shown analytically by \citet{2022arXiv220105589R}, the latter effect from the dark matter wake itself on the response of stars or satellites is a minor one. However, these additional effects will all be included in a self-consistent manner in the numerical simulations presented and analysed from Sect.~\ref{sec:Sim} onwards.

In Fig. \ref{fig:shifts} we reproduce the COM shifts (as obtained from figures 4 and 6 in \citetalias{GC21}). We subtract the respective distance-dependent values from the Cartesian positions and velocities of the observed MW satellites, for which we adopt the dataset by \citet{2022A&A...657A..54B} who provide proper motions based on the early-third data release (EDR3) of \textit{Gaia}. Here and in the following we focus on satellites in the Galactocentric distance range of 50 to 250\,kpc, which results in a sample of 31 dwarfs. For more nearby objects no effect due to the LMC is expected according to \citetalias{GC21}, while more distant objects have only weakly constrained orbital poles due to the large proper motion uncertainties.

Panel b of Fig. \ref{fig:shifts} plots the most-likely positions and velocities of the MW satellites. Also shown are the shifted positions and velocities. For most dwarfs the shifts are minor. The overall orbital directions do not appear to be substantially altered; especially the objects co-orbiting closely along the VPOS (red) keep their preferred orbital sense (clockwise).  

This impression is confirmed by Fig. \ref{fig:shifts2}, which shows the relative change in position $\mathbf{r}$ and velocity $\mathbf{v}$, the relative change in the norm of the specific angular momentum $h$, and the difference $\Delta \theta_\mathrm{pole}$\ in the direction of the resulting orbital pole (direction of the angular momentum) before and after applying the shifts of each dwarf galaxy, respectively. The positions shift by no more than 10\% of the Galactocentric distance of each dwarf, most velocities change by no more than 20\%, though for some the velocity change can reach up to 50\% of their full 3D velocity. Note, however, that a complete reversal of the velocity requires a change of 200\%. The specific angular momenta change by no more than 30\%. Most importantly, the direction of the orbital poles of the majority of MW dwarfs changes by no more than $\sim 5^\circ$. Only 7 of the 31 considered dwarfs display $\Delta \theta_\mathrm{pole} > 10^\circ$. Of these, only one is strongly co-orbiting and one strongly counter-orbiting relative to the VPOS. The mean (median) change on orbital pole direction for the MW dwarfs is $3^\circ (6^\circ)$.

\begin{figure*}
\gridline{
\fig{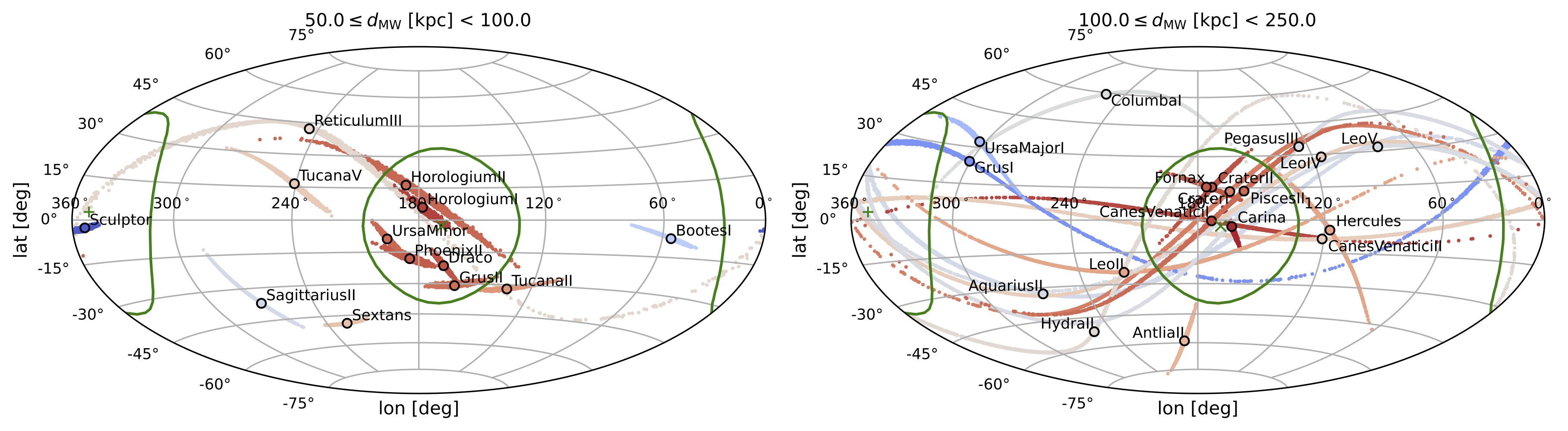}{0.99\textwidth}{(a) Observed orbital poles}
}
\gridline{
\fig{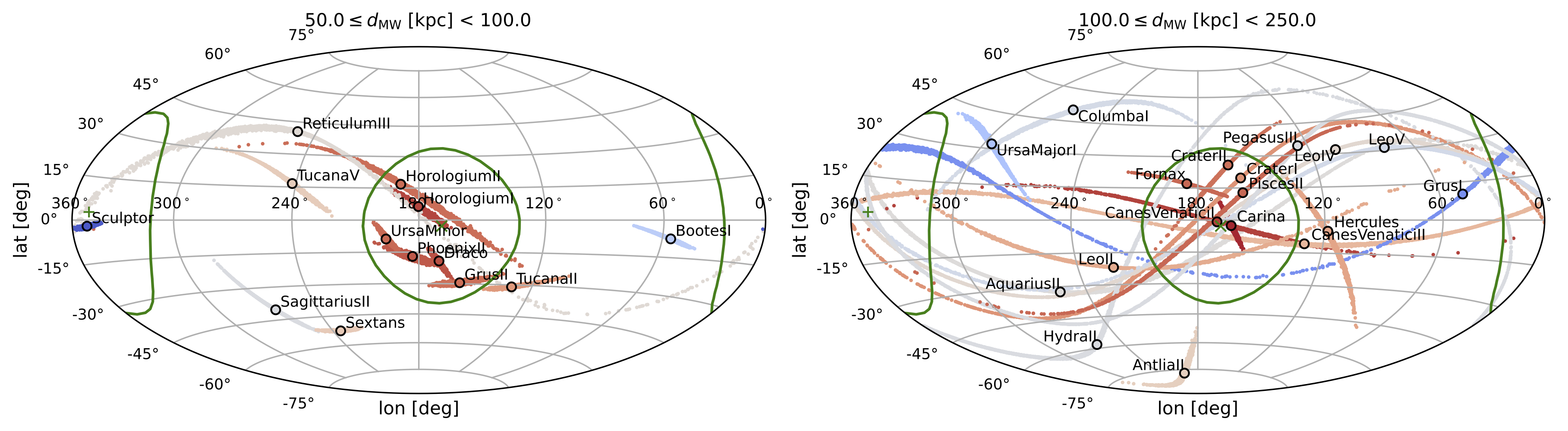}{0.99\textwidth}{(b) Orbital poles after correcting for LMC-induced reflex motion and COM shift}
}
\caption{Directions of orbital poles for the observed MW satellites calculated relative to the MW center (upper panel a), and after shifting the dwarf galaxy positions and velocities as in Fig. \ref{fig:shifts} to correct for the effect of the LMC (lower panel b). The most-likely orbital poles are shown as open circles, and the orbital poles from 5000 Monte-Carlo realizations drawing from the measurement uncertainties as dots. The colors follow those in Fig. \ref{fig:shifts}, with strongly co-orbiting dwarfs in red and counter-orbiting ones in green. If the LMC's influence were the main reason for the observed overdensity of orbital poles towards the VPOS normal vector (green cross), then the lower panels should show a much reduced overdensity. They do not. Most orbital pole directions are barely affected by the shifts. 
\label{fig:Poles}}
\end{figure*}

The impact on the distribution of orbital poles is illustrated in Fig. \ref{fig:Poles}.
To account for measurement uncertainties in distances, velocities, and proper motions, we draw 5000 realizations for each dwarf.
The distributions without and with shifts are virtually indistinguishable. In both cases a clear overdensity of orbital poles close to the center of the figures is apparent.
Overall, the number of MW dwarfs with most likely orbital poles within $36.87^\circ$\ (10\% of the area of the sphere around the VPOS normal direction, see e.g. \citealt{2018A&A...619A.103F, 2021ApJ...916....8L}) is 12 for the observed orbital poles out of the 31 dwarfs (ignoring two counter-orbiting dwarfs)\footnote{Note that many of the others remain consistent with aligning with the VPOS within their considerable proper motion uncertainties; their orbital poles are only very weakly constrained.}. This corresponds to an orbital pole enhancement in the 10\% VPOS region of almost 300\% over the expected isotropic share of 3.1 orbital poles out of 31 dwarfs. After applying the shifts, all 12 orbital poles remain within the VPOS region, and the median alignment angle $\theta_\mathrm{VPOS}$\ with the VPOS changes from $59^\circ$ to $61^\circ$. 
 
It has been demonstrated that the LMC has likely brought along a number of dwarf galaxies as satellites of its own \citep{2020MNRAS.495.2554E, 2020ApJ...893..121P, 2022A&A...657A..54B}. These would not constitute independent objects, and could boost the VPOS orbital enhancement given that they should follow orbits similar to the LMC. The exact strength of the observed orbital pole enhancement thus depends on which dwarfs were satellites of the LMC before infall. By considering only objects beyond 50\,kpc in this work, we already ensure that many likely LMC satellites, specifically Carina\,II, Carina\,III, Hydrus\,I and Reticulum\,II, are not part of our sample. However, of the likely LMC satellites identified by \citet{2022A&A...657A..54B}, Phoenix\,II and Horologium\,I are part of our sample of 31 dwarfs. In addition, \citet{2022A&A...657A..54B} report inconclusive findings on the LMC-association of Horologium\,II, which we thus consider as a potential past LMC satellite. Note that we must not exclude dwarf galaxies that had a recent interaction but were unlikly to have been brought in as LMC satellites, because such interactions with the LMC are exactly what is being studied here and in particular in the following simulation-based investigation. An example for this is Grus\,II, which \citet{2022A&A...657A..54B} report to not originate with the Magellanic system, but to merely have interacted with the LMC recently. We furthermore do not consider the classical MW satellites Carina or Fornax as likely past LMC satellites, because both were excluded as such by \citet{2020ApJ...893..121P}, and also \citet{2022A&A...657A..54B} report this as unlikely. Excluding the two likely LMC satellites with orbital poles within the VPOS region from our sample of 31 dwarfs results in an orbital pole enhancement of 250\%, while also excluding the possible past LMC satellite Grus\,II results in an enhancement of 220\%. The enhancement thus remains substantial even if accounting for dwarfs brought in as LMC satellites.

We note that this is only a first-order estimate of the influence of the LMC. More accurate results require full orbit modelling under the influence of the time-varying potential sourced by the MW and LMC. Recently, \citet{2021arXiv211000018C} have presented such a work. They report that re-winding the satellite orbits under the influence of a massive LMC, and then forward-integrating them again without an LMC until present time, results in no significant change in the orbital pole distribution and preserves its strong non-uniformity. We are thus confident that the LMC's impact on the orbital pole clustering observed for the MW satellites is indeed very minor.

This begs the question as to why the model of \citetalias{GC21} appears to be insufficient to account for the observed orbital pole clustering. Our hypothesis is that the reason lies in different orbital properties between typical dark matter particles in a halo, on which their work is based, and the MW dwarf galaxies. \citetalias{GC21} used a slightly radially biased setup for their dark matter halo particle velocities. If an orbit is strongly radial, then a minor shift in position or velocity of the reference point for angular momentum calculations has much greater impact on the orbital pole direction than if the orbit were more circular. While dark matter particles can be on highly radial orbits, the observed MW dwarf galaxies have been found to follow more tangentially biased orbits \citep{ 2017MNRAS.468L..41C,2019MNRAS.486.2679R,2021arXiv210911557H}. 
This is not entirely unexpected in a $\Lambda$CDM framework. \citet{2004MNRAS.352..535D} report for galaxy cluster mass simulations that dark matter particles in the inner regions of halos are on slightly more radial orbits than subhalos, and that subhalos typically have higher velocities than dark matter particles.
The presence of a baryonic central galaxy seems to further strengthen this difference, because the additional potential in the inner region of a halo results in stronger tidal forces that even more efficiently destroy subhalos on radial orbits. Using the hydrodynamical FIRE simulations, \citet{2017MNRAS.471.1709G} showed that this leads to more tangentially biased subhalo velocities when compared to dark-matter-only simulations. They also showed that the surviving subhalos in the hydrodynamical runs have considerably higher tangential velocities within 100\,kpc than those in dark-matter-only runs, which implies higher angular momenta. 
This effect has been confirmed by \citet{2019MNRAS.487.4409K} for the PhatELVIS simulations which contain an analytically grown MW-like disk potential, and by \citet{2019MNRAS.486.2679R} who find that satellites in the Auriga simulations are on more tangentially biased orbits, probably because the Auriga halos contain rather massive baryonic disks that destroy subhalos on radial orbits.

\section{Our Simulation} \label{sec:Sim}

To test our hypothesis, we run our own simulation of a MW-LMC-like interaction. The $N$-body simulation is run with \textsc{Gyrfalcon} \citep{2000ApJ...536L..39D}, and we use \textsc{Agama} \citep{2019MNRAS.482.1525V} to generate initial conditions. For the MW, we use the Model 1 of \citet{2008gady.book.....B} as given in \textsc{Agama}, which includes a stellar bulge, a thin disk, a thick disk, and a dark matter (DM) halo. We make two changes: (i) we use the same halo mass of $1.2\times10^{12}$ M$_\odot$ as \citetalias{GC21}, and (ii) in order to have a spherical halo, we remove the z-axis flattening coefficient of its potential. We generate $2\times10^5$ particles for the stars and $8\times10^5$ for the dark matter halo. For the LMC, we generate $10^5$ particles based on the spheroid potential of \textsc{Agama} for a total mass of $1.8\times10^{11}$ M$_\odot$, following the fiducial model of \citetalias{GC21}. 
The initial conditions of our simulated MW halo model deliberately adopt isotropic orbits, in order to facilitate our aim to examine the effect of the infall of the LMC model on the halo particles as a function of their specific angular momenta. The anisotropy parameter calculated using all dark matter halo particles thus is very close to zero: $\beta  = 1 - \frac{\sigma_\theta^2 + \sigma_\phi^2}{2 \sigma_r^2} = 0.0083$, where $\sigma$\ are the velocity dispersions of the particles in the three spherical coordinate components $(r, \theta, \phi)$.

Initially we place the LMC at $(x, y, z) = (12, 215, 130)$\,kpc and $(v_x, v_y, v_z) = (12, 13, -77)\,\mathrm{km\,s}^{-1}$ in the MW-centred reference frame, a slight alteration of the initial conditions of \citetalias{GC21}. We run the simulation for 2 Gyr. For each snapshot, we use the {\it snapcenter} function of \textsc{Nemo} \citep{1995ASPC...77..398T} to obtain the positions and velocities of the COM of the MW based on its stars, then we shift the positions and velocities of all particles by these values in order to make the MW the center of phase space. 
In this reference frame, the coordinates of the COM of the LMC after 2 Gyr are $(x, y, z) = (-1.87, -31.42, -17.07)$\,kpc and $(v_x, v_y, v_z) = (-48.61, -207.98, 150.63)\,\mathrm{km\,s}^{-1}$. For comparison, the present day values retained by \citet{2021MNRAS.501.2279V} are $(x, y, z) = (-0.6, -41.3, -27.1)$\,kpc and $(v_x, v_y, v_z) = (-63.9, -213.8, 206.6)\,\mathrm{km\,s}^{-1}$.

\section{Results} \label{sec:AngMom}

\begin{figure*}
\gridline{
\fig{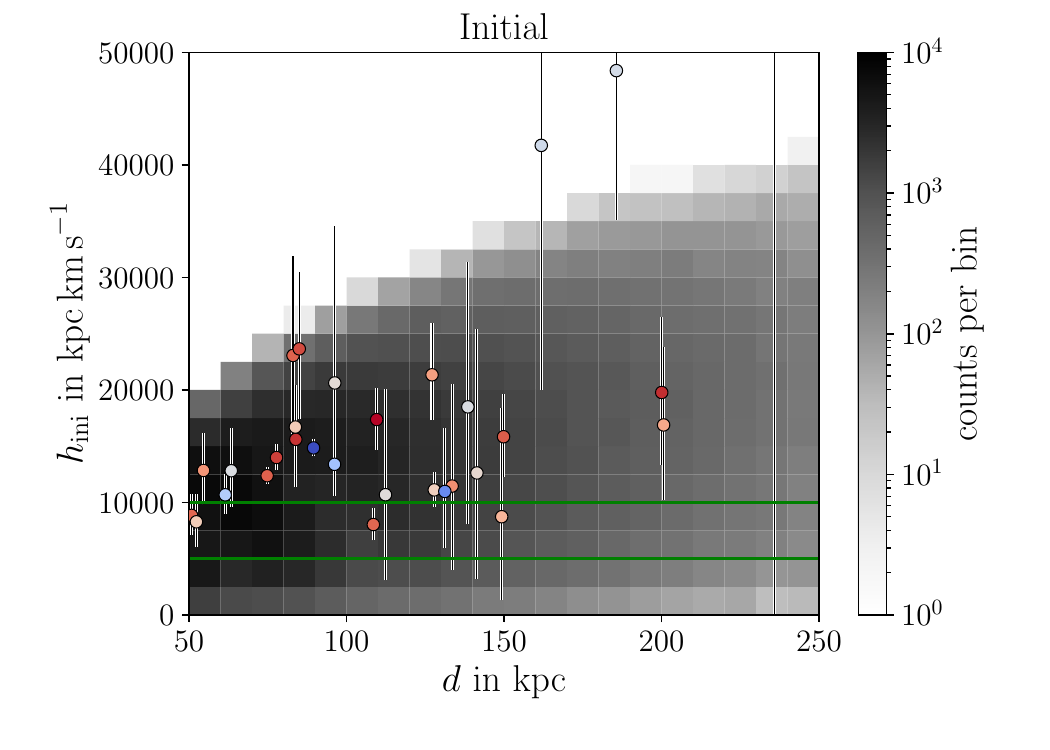}{0.5\textwidth}{(a): Initial angular momentum}
\fig{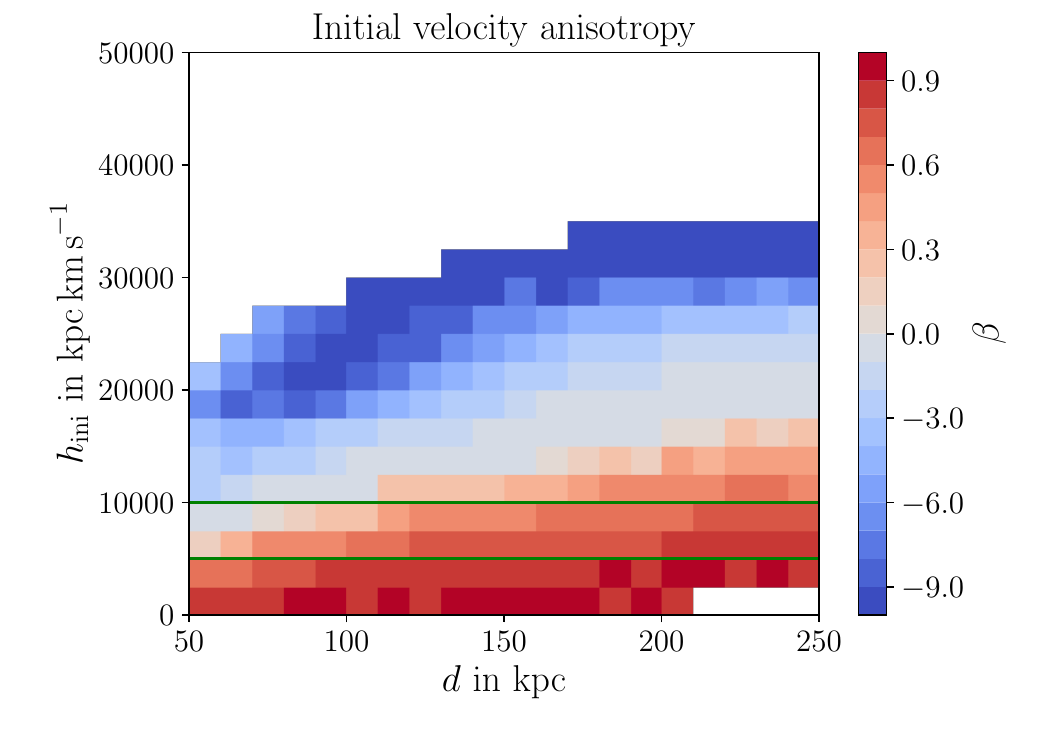}{0.5\textwidth}{(b): Anisotropy parameter for bins in panel (a)}
}
\gridline{
\fig{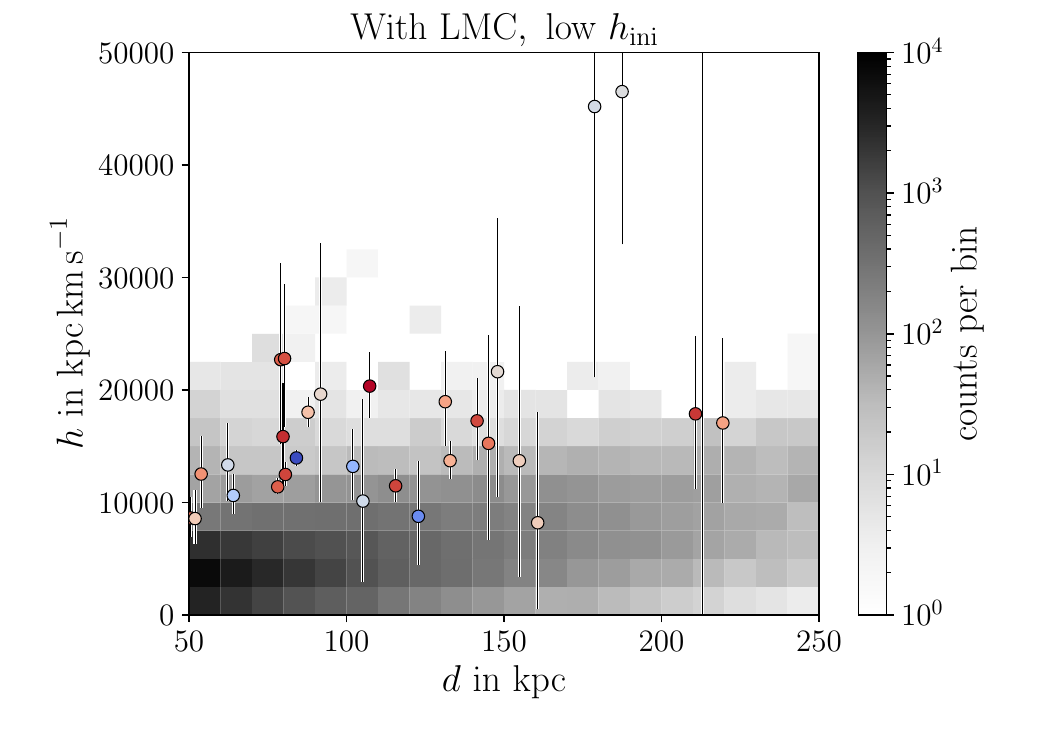}{0.333\textwidth}{(c): Low initial angular momentum}
\fig{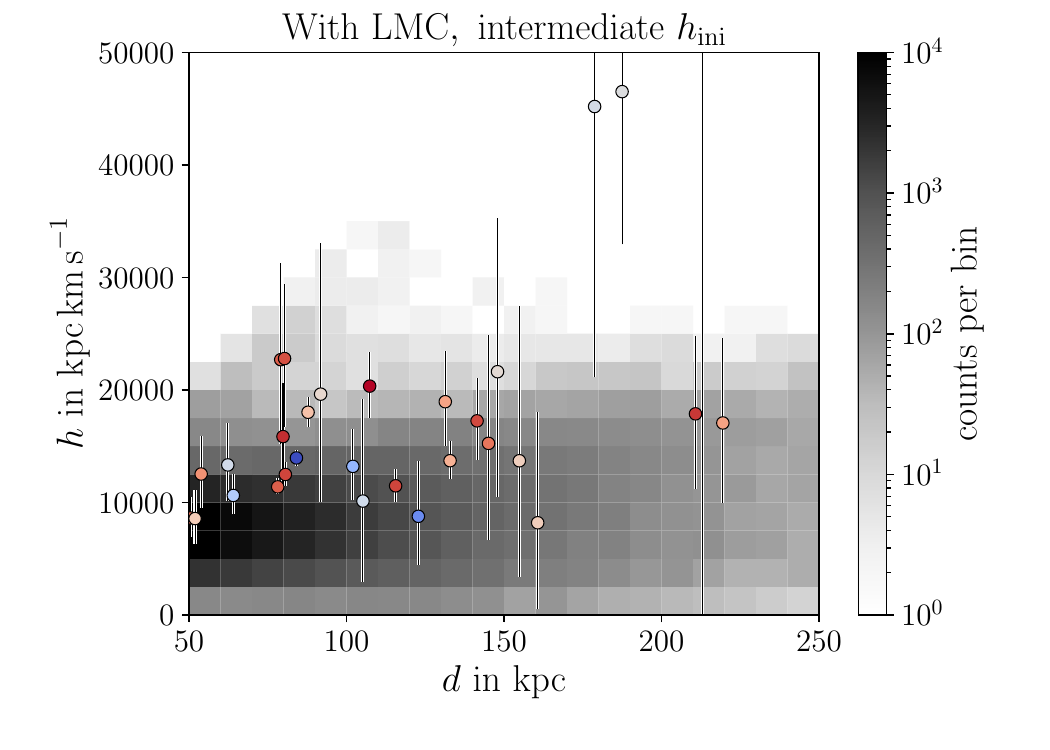}{0.333\textwidth}{(d): Intermediate initial angular momentum}
\fig{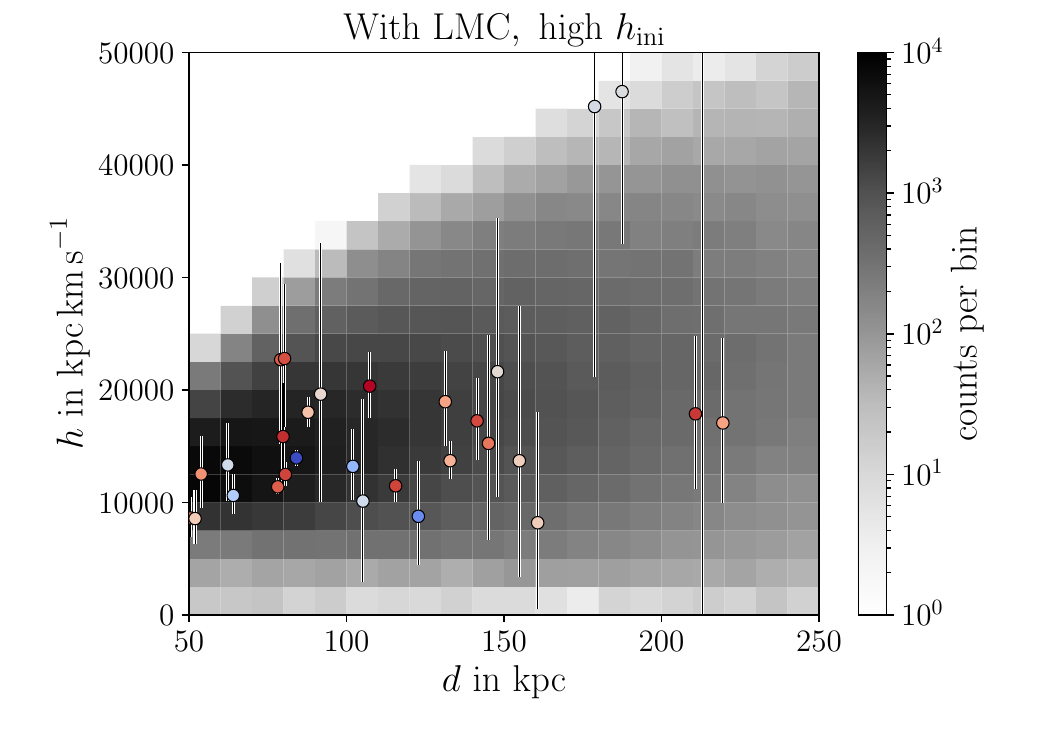}{0.333\textwidth}{(e): High initial angular momentum}
}
\caption{Specific angular momenta of MW dwarfs (symbols) and dark matter particles in our simulation (density maps, with logarithmic scale). Panel (a) shows the initial specific angular momentum $h_\mathrm{ini}$\ (and the one inferred after shifting the observed MW dwarf positions and velocities), while panel (b) shows the velocity anisotropy parameter $\beta$\ for all particles contributing to the corresponding bins in panel (a).
Panels (c) to (e) show the angular momentum $h$\ of the same set of particles in the final simulation snapshot (and the angular momenta of the MW satellites relative to today's MW center), split into low, intermediate, and high initial angular momentum bins (as indicated by the green lines in panel a). Particles are selected from the distance range of 50 to 250\,kpc in the final snapshot. The observed MW dwarfs best agree with the simulation particles in the high angular momentum bin.
\label{fig:AngMom}}
\end{figure*}

We follow the evolution of individual DM particles in the simulation from the initial to final snapshot, selecting all particles at a distance of 50 to 250\,kpc from the MW analog center in the final snapshot. We divide them in three bins of low ($h_\mathrm{ini} < 5 \times 10^3\,\mathrm{kpc\,km\,s}^{-1}$, $N = 49,884$), intermediate ($5 \times 10^3\,\leq h_\mathrm{ini} < 10^4\,\mathrm{kpc\,km\,s}^{-1}$, $N = 109,905$) and high ($10^4\,\mathrm{kpc\,km\,s}^{-1} \leq h_\mathrm{ini}$, $N = 232,533$) initial specific angular momenta. While the two bins with lower angular momentum constitute only 41\% of all particles in the considered distance range, we will show in the following that they contribute the bulk to the orbital pole overdensity around the VPOS direction.

Fig. \ref{fig:AngMom} plots the specific angular momenta of the observed MW dwarfs against their Galactocentric distance. It compares to particles in our simulation, selected to reside in the considered distance range {\it at the end of the simulation}. 
The LMC affects the angular momenta of the particles, spreading them beyond their initial bin boundaries, though overall they preserve their ranking in specific angular momentum (note the density scale is logarithmic). 
The observed MW dwarfs agree best with the final particle distribution of the high-angular-momentum bin. As expected, the MW dwarfs have relatively high specific angular momenta compared to dark matter particles in a MW-like halo.
To more quantitatively judge this visual impression, we calculate the likelihood ratios of the high angular momentum bin to the low and the intermediate angular momentum bins, respectively. 

\begin{figure*}
\gridline{
\fig{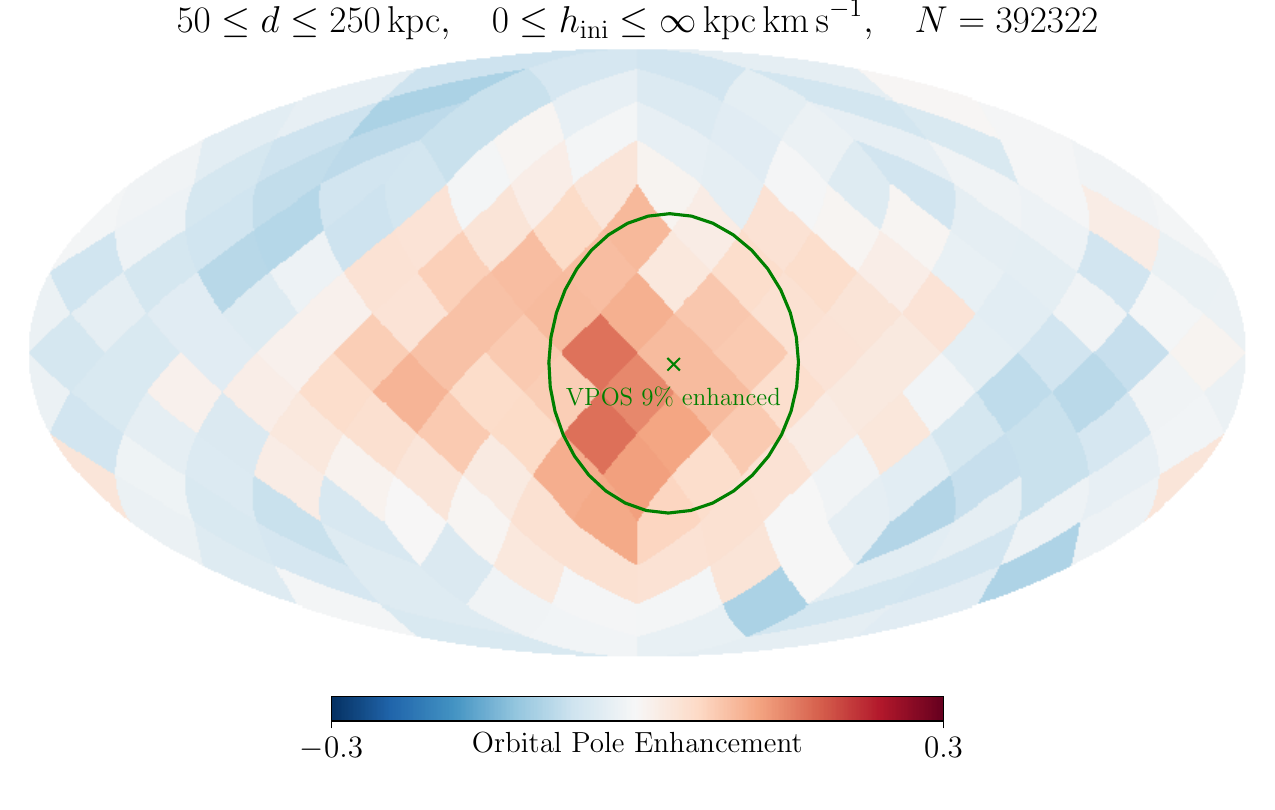}{0.5\textwidth}{(a) all particles in distance range}
\fig{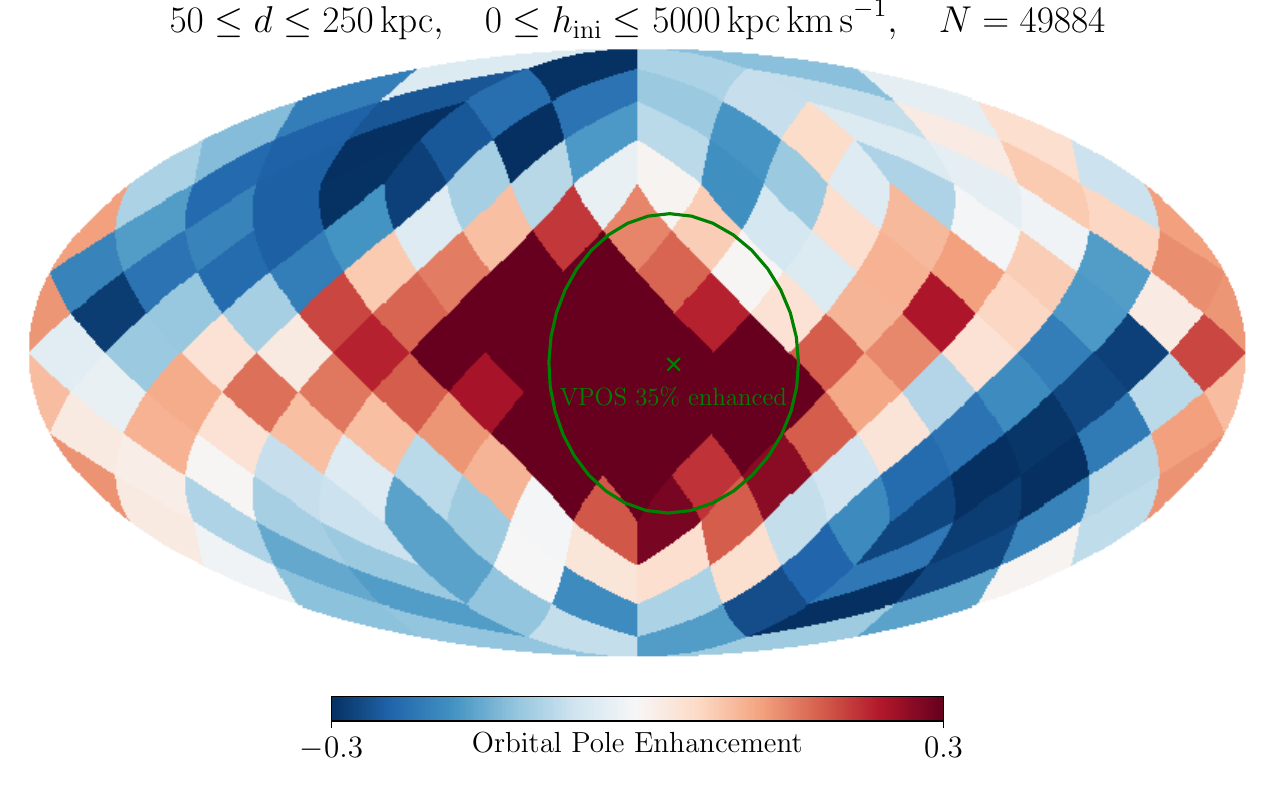}{0.5\textwidth}{(b) low initial angular momentum}
}
\gridline{
\fig{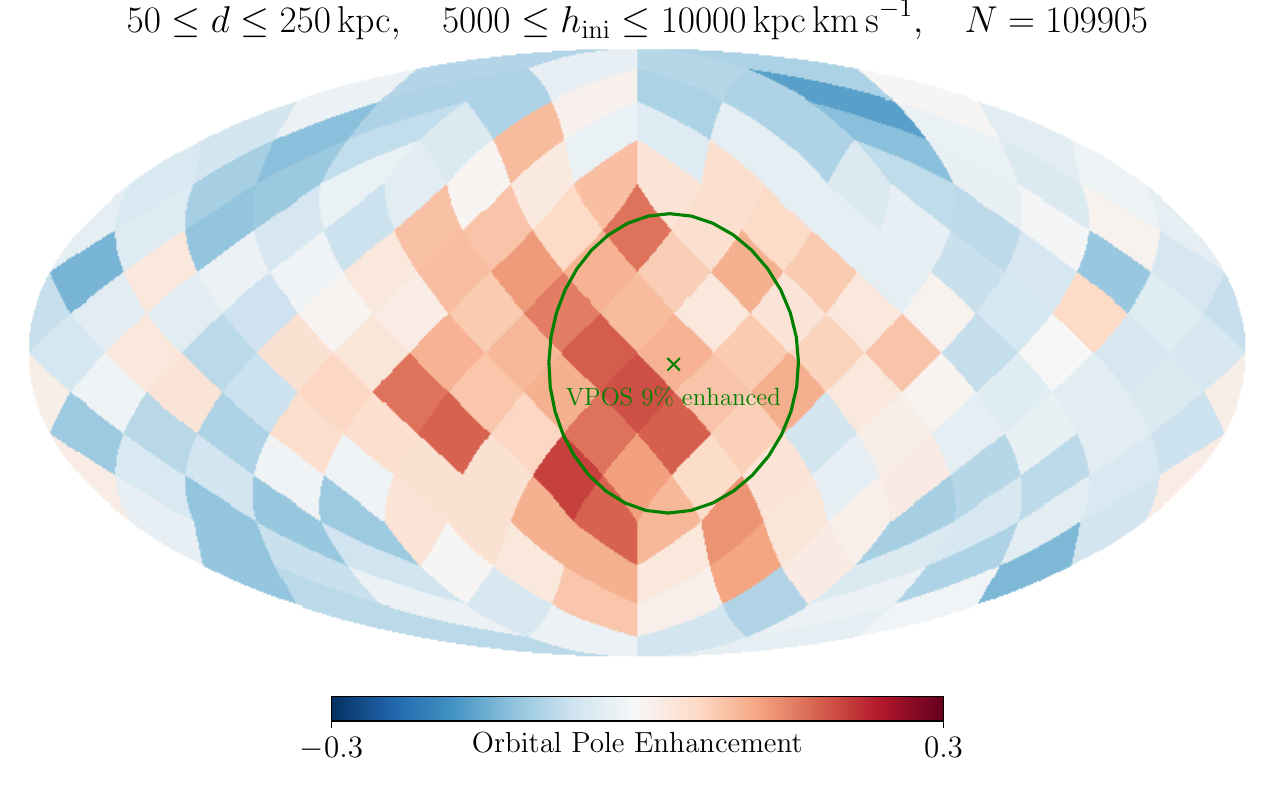}{0.5\textwidth}{(c) intermediate initial angular momentum}
\fig{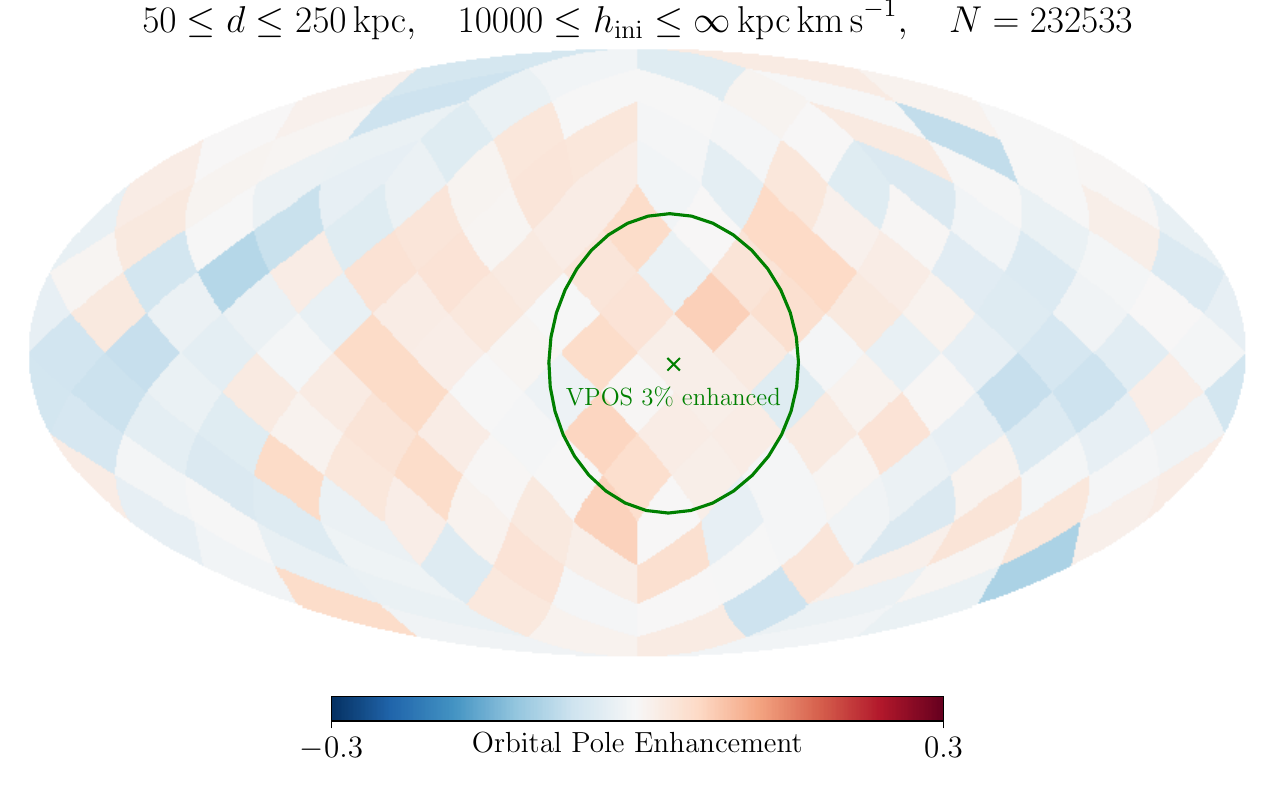}{0.5\textwidth}{(d) high initial angular momentum}
}
\caption{All-sky maps in Galactic Coordinates of the orbital pole enhancement for dark matter particles in our simulation. Panel (a) contains all particles within a distance range of 50 to 250\,kpc, while the other panels only show subsets based on the particles' initial specific angular momenta. The green cross marks the VPOS normal vector direction, and the green circle indicates the 10\% area on the sphere where the relative VPOS enhancement is measured. While panel (a) confirms the overall findings of \citetalias{GC21} that an LMC-like infall enhances the density of orbital poles towards the VPOS normal, low angular momentum particles (panel b) are most affected and show a pronounced overdensity, while high angular momentum particles (panel d) -- which are comparable to the observed MW dwarfs -- show only a very slight overdensity.
\label{fig:LMCeffectObs}}
\end{figure*}

The likelihoods are estimated using the maps of angular momenta and distances in Fig. \ref{fig:AngMom} as follows: for each observed satellite, we identify the Galactocentric distance $d$\ bin (of 10\,kpc width) it falls in. We then calculate the weighted contribution of this observed satellite's angular momentum to each of the angular momentum bins (of 2500\,kpc\,km\,s$^{-1}$\ width) at this $d$. This is done by adopting the most-likely measured value of $h$\ for this satellite and assuming a normal distribution with a width as given by the uncertainties on $h$ (see error bars in Fig. \ref{fig:AngMom}). Since $h$\ can only be positive we cut off these distributions at $h = 0$\ and re-normalize them accordingly. To avoid being dominated by a few high-$h$\ outliers, we also do not consider satellites or particles with $h > 5 \times 10^4\,\mathrm{kpc\,km\,s}^{-1}$. The likelihood for one individual satellite is then the weighted contribution of simulation particles in these bins, normalized to the total number of simulation particles in the considered set (i.e. either low, intermediate, or high $h_\mathrm{ini}$). Finally, the individual likelihoods of all observed satellites are multiplied to obtain the overall likelihood. We find that the likelihood ratio of high-to-low $h_\mathrm{ini}$\ is $1.3 \times 10^{22}$, while that of high-to-intermediate $h_\mathrm{ini}$\ is $4.1 \times 10^{8}$. This strongly confirms the impression that the set of observed Milky Way satellite galaxies match best with the high initial angular momentum bin.

To test whether the particles most affected by the LMC-effect indeed have more radial, eccentric orbits, we have also calculated the anisotropy parameter $\beta$\ for bins in initial angular momentum $h_\mathrm{ini}$\ and distance $d$, see panel (b) in Fig. \ref{fig:AngMom}. As expected, the high-angular momentum particles are dominated by tangentially biased orbits ($\beta < 0$, blue in the plot), while the low- and intermediate-angular momentum particles are characterized by radially biased orbits ($\beta > 0$, red in the plot).

\subsection{Orbital Pole Enhancement}

\begin{figure*}[ht!]
\gridline{
\fig{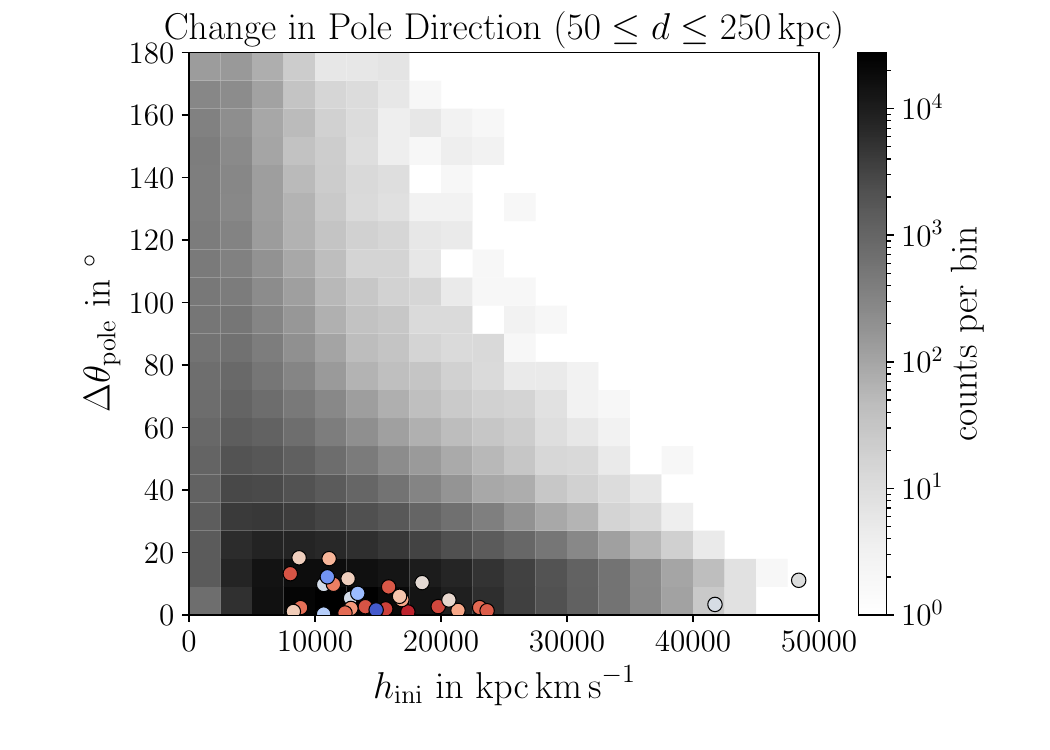}{0.5\textwidth}{(a)}
\fig{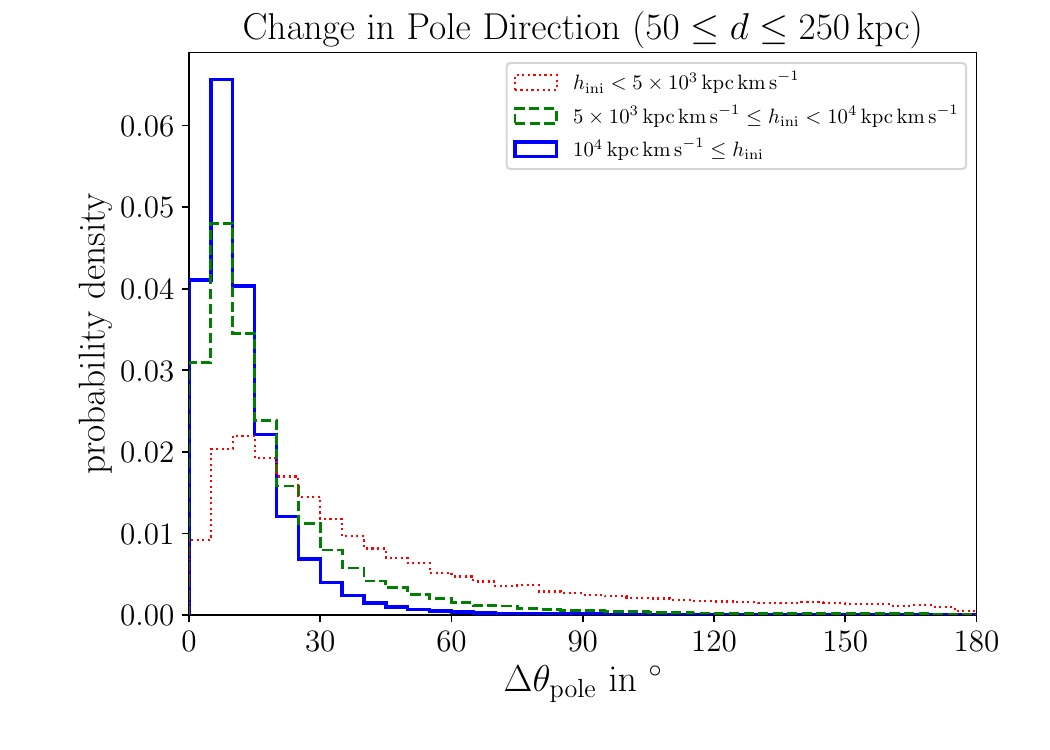}{0.5\textwidth}{(b)}
}
\caption{ The change in orbital pole direction of particles in the simulation depends on the specific angular momentum they had initially. 
Panel (a) shows that particles with smaller initial specific angular momentum $h_\mathrm{ini}$\ show larger changes in the direction of their orbital poles as measured by the angle $\Delta \theta_\mathrm{pole}$\ between their orbital pole in the initial and final simulation snapshot. Also shown are the observed MW dwarfs.
Panel (b) shows the distribution of $\Delta \theta_\mathrm{pole}$\ for three specific angular momentum bins. Low angular momentum particles (red dotted line) are affected most, while high angular momentum particles (blue solid line) show the least change in their orbital pole directions.
 \label{fig:polechange}}
\end{figure*}

The enhancement of the density of orbital pole directions for the particles in our simulation are plotted in Fig. \ref{fig:LMCeffectObs}. We follow the method of \citetalias{GC21}, and calculate for each bin in the Healpix map how many orbital poles in the particle sample contribute to it, subtract the expected number for an isotropic distribution, and divide by the latter. Despite the lower resolution of the maps due to our smaller particle numbers, we clearly confirm the results of \citetalias{GC21}. The infall of an LMC-like object onto a MW-like host results in an enhancement of orbital poles of dark matter particles in the general direction of the VPOS.

We also calculate how many more orbital poles than in an isotropic distribution contribute to the VPOS region. We assume the VPOS normal points to Galactic coordinates $(l, b)_\mathrm{VPOS} = (169.3^\circ, -2.8^\circ)$\ to be consistent with previous works \citep[e.g.][]{2018A&A...619A.103F, 2021ApJ...916....8L,GC21}. Also following these previous works, we consider a region of $36.87^\circ$\ around this direction, which corresponds to 10\% of the area on the sphere. The excess of the number of particle orbital poles in this area in the final simulation snapshot over the corresponding number using the same set of particles in the initial setup, divided by the latter, is the relative VPOS enhancement. Within the adopted distance range of 50 to 250\,kpc, we find an overall VPOS enhancement of 9\%. This means that there are 1.09 times the number of orbital poles within the VPOS region than expected from isotropy. As the region constitutes 10\% of the sphere, for 31 dwarf galaxies 3 should typically be found in the region if they are isotropically distributed. The LMC effect would then suggest an enhancement of about one quarter of an additional orbital pole. This is hardly sufficient to account for the observed clustering of 12 (or more within their uncertainties) out of 31 orbital poles aligned with the VPOS.

In the other panels of Fig. \ref{fig:LMCeffectObs} we separate the DM particles by their initial angular momentum. The resulting maps of orbital pole enhancement strongly confirm our suspicion that mainly particles with small angular momenta are affected by the LMC's influence. The low angular momentum bin shows the strongest VPOS enhancement of 35\%. The pole density distribution also displays a slight s-shape that further confirms the results reported by \citetalias{GC21}. Particles with intermediate specific angular momenta still display some enhancement in the orbital pole density, but to a much reduced degree (VPOS enhancement of 9\%). Once only high angular momentum particles are considered, the map mainly consists of noise, with no substantial enhancement of orbital poles towards the VPOS direction (VPOS enhancement of 3\%).

\subsection{Change in Orbital Pole Directions}

We now investigate how the orbital poles change between the initial and the final simulation snapshot to assess how much the LMC-like infall has affected the orbital direction of each individual particle. 
Panel (a) of Fig. \ref{fig:polechange} plots the angle between the initial and the final direction of orbital poles, $\Delta \theta_\mathrm{pole}$. The vast majority of simulation particles (note the logarithmic color scale) overlap with the region covered by the observed MW dwarfs. There is a clear tendency for particles with higher initial specific angular momentum to have smaller changes in their orbital pole direction. 
Panel (b) of Fig. \ref{fig:polechange} confirms this. Low angular momentum particles display an extended tail to high $\Delta \theta_\mathrm{pole}$, some reaching as high as $\Delta \theta_\mathrm{pole} = 170^\circ$\ and thus almost flip their orbital direction. Their mean (median) change in orbital pole direction is $43.9^\circ (29.2^\circ)$. The particles with intermediate angular momentum show smaller changes of $19.7^\circ (12.8^\circ)$. The vast majority of particles with high initial angular momentum, however, change their orbital pole direction by less than $20^\circ$, and correspondingly their mean (median) $\Delta \theta_\mathrm{pole}$\ are only $12.5^\circ (9.4^\circ)$.

\subsection{Change in Alignment with VPOS}

\begin{figure*}[ht!]
\gridline{
\fig{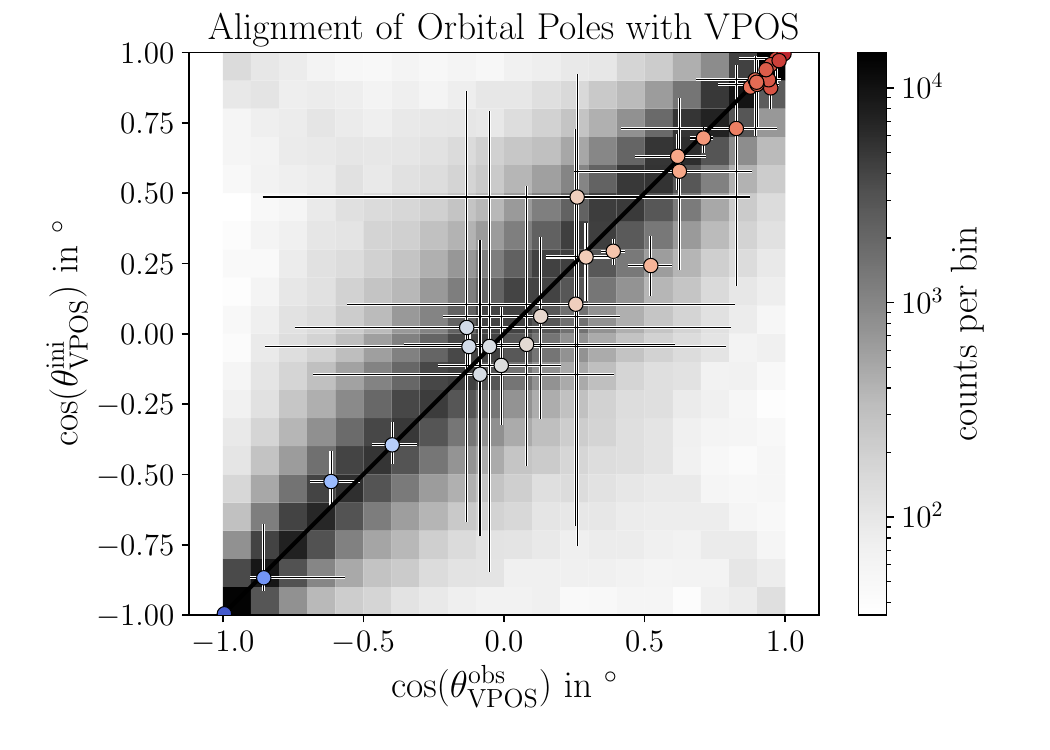}{0.5\textwidth}{(a)}
\fig{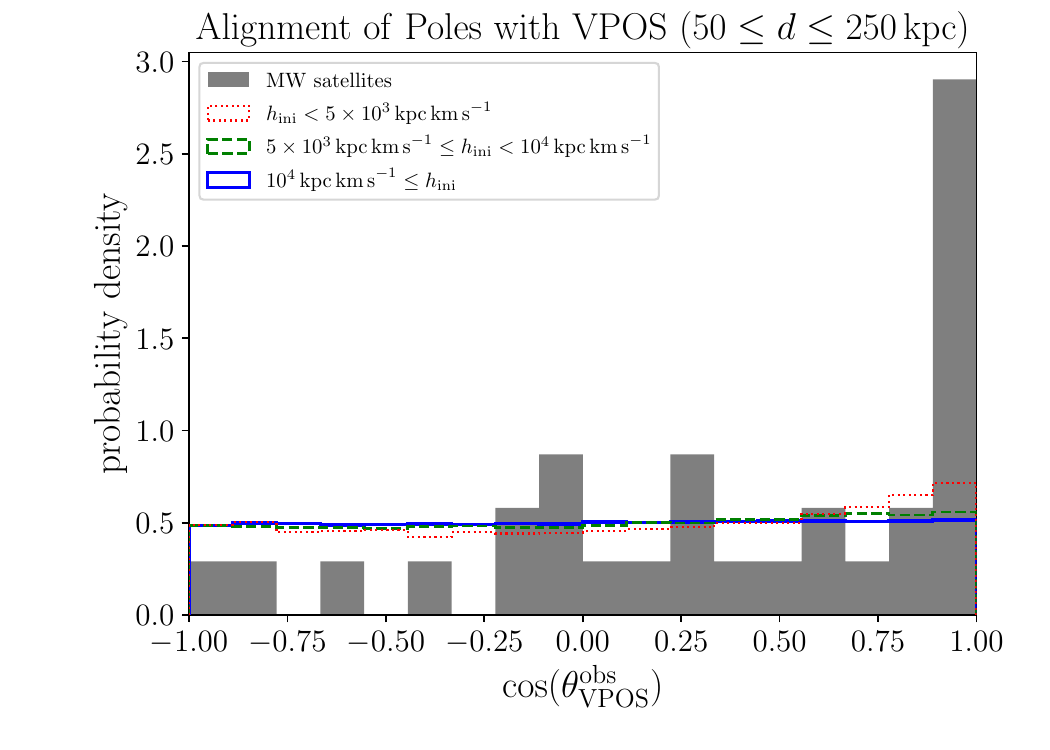}{0.5\textwidth}{(b)}
}
\caption{ The enhancement in the alignment of orbital poles with the VPOS normal direction of particles in the simulation depends on the specific angular momentum they had initially. 
Panel (a) plots the Cosines of the angle between the VPOS normal vector and the particle orbital poles as measured in the initial ($\theta_\mathrm{VPOS}^\mathrm{ini}$) and final ($\theta_\mathrm{VPOS}^\mathrm{obs}$) simulation snapshot. The corresponding angles for the observed (and shifted) MW dwarf positions and velocities are also shown as colored symbols.
Panel (b) compares histograms of $\theta_\mathrm{VPOS}^\mathrm{obs}$\ for three different bins in specific angular momentum for the particles in the simulation, with the alignment of orbital poles of the observed set of MW dwarfs. Only the low angular momentum (red dotted line) particles show some enhancement towards the VPOS normal (at $\cos(\theta_\mathrm{VPOS}^\mathrm{obs}) = 1$). The high angular momentum particles (blue solid line) follow a flat distribution as expected for isotropically distributed orbital poles.  The observed MW dwarfs have specific angular momenta that are comparable to the high angular momentum bin, but they display a very pronounced overdensity of orbital poles close to the VPOS normal, which is more than an order of magnitude higher than even the enhancement shown by the low angular momentum particles.
 \label{fig:VPOSangle}}
\end{figure*}

Changing the direction of the orbital poles does not necessarily imply an enhancement of poles towards the VPOS. To quantify this, we measure the angle $\theta_\mathrm{VPOS}$\ between each particle's orbital pole and the normal vector to the VPOS. We again do this for both the initial snapshot which should be comparable to the ``corrected'' positions after shifting the observed MW dwarf positions and velocities ($\theta_\mathrm{VPOS}^\mathrm{ini}$), and the final one which is comparable to the observed situation ($\theta_\mathrm{VPOS}^\mathrm{obs}$). Panel (a) of Fig. \ref{fig:VPOSangle} compares these two quantities\footnote{The Cosines are plotted as they ensure that isotropically distributed directions result in a flat distribution.}. If the clustering of observed MW orbital poles close to the VPOS were substantially affected by the LMC infall, $\theta_\mathrm{VPOS}^\mathrm{ini}$\ should show a much wider distribution than $\theta_\mathrm{VPOS}^\mathrm{obs}$, and in particular the dwarfs with closely aligned poles (red) should scatter appreciably towards lower $\cos(\theta_\mathrm{VPOS}^\mathrm{ini})$. Yet, the distributions largely follow the diagonal in the figure, suggesting that the overall changes are small.

The situation is even clearer when looking at histograms of $\theta_\mathrm{VPOS}^\mathrm{obs}$\ (panel (b) in Fig. \ref{fig:VPOSangle}). 
An isotropic distribution of directions on the sphere, such as the DM particle orbital poles in our initial simulation snapshot, gives a flat histrogram with probability density 0.5. Clearly, the simulation particles in the final snapshot mostly follow this distribution, in particular for the high angular momentum bin. While an isotropic distribution should have a mean (median) alignment angle of $90^\circ$, the low, intermediate, and high specific angular momentum bins show only mildly smaller angles, namely $85.8^\circ (84.6^\circ)$, $88.0^\circ (87.2^\circ)$, and $89.4^\circ (89.1^\circ)$ respectively. The corresponding values for the observed MW dwarfs are $63.0^\circ (58.5^\circ)$. Only the low angular momentum bin shows some clear excess towards $\cos(\theta_\mathrm{VPOS}^\mathrm{obs}) = 1$, i.e. around the VPOS direction. However, this excess is neither comparable in strength to the distribution of observed dwarf galaxy orbital poles, nor do the observed MW satellites have such low specific angular momenta.

\subsection{Dependency on Distance from Host}

\begin{figure}[ht!]
\gridline{
\fig{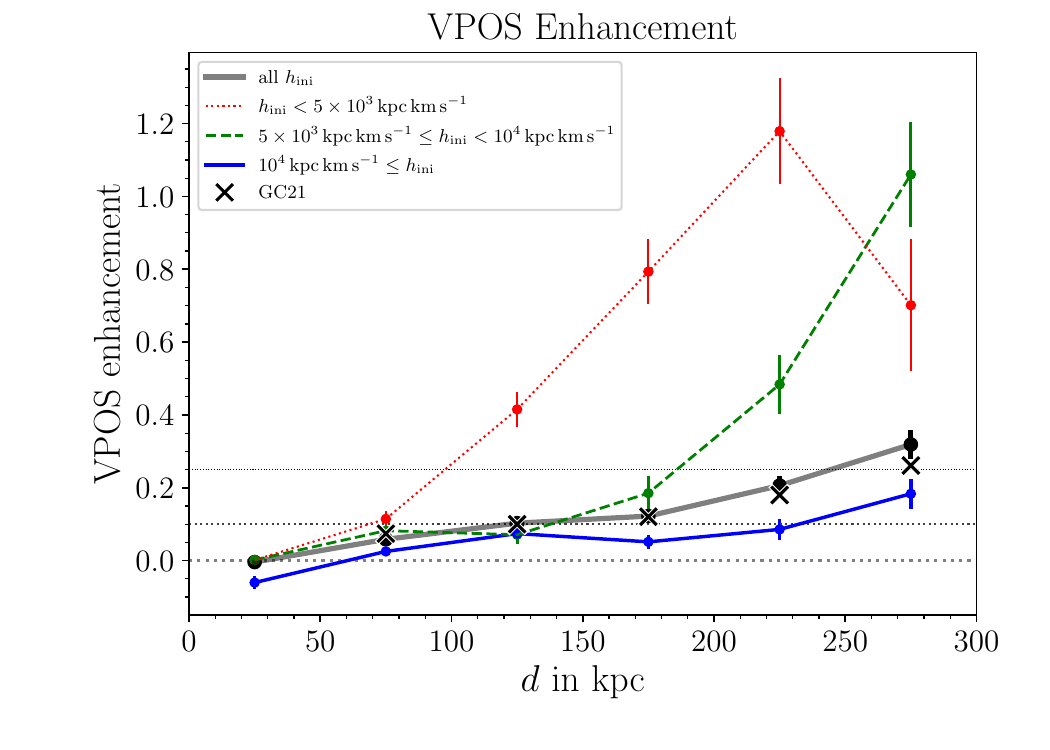}{0.48\textwidth}{}
}
\caption{The enhancement of orbital pole density of dark matter particles in the simulation over that expected for isotropy in the 10\% region around the VPOS normal vector, considering only particles with Galactic latitude $| l | > 20^\circ$. The enhancement is measured as a function of distance from the host center $d$, in radial shells of 50\,kpc width. While low angular momentum particles can show very pronounced enhancements at larger distances, the higher-angular momentum particles, which are more comparable to the observed MW satellites, show a substantially lower enhancement and also dominate the sample of particles at larger distances. Also plotted are the VPOS enhancements as reported in \citetalias{GC21} and extracted from their figure 15. Dotted grey lines of decreasing width indicate enhancements of 0, 10, and 25\%. \label{fig:VPOSenhancement}}
\end{figure}

These results are not strongly dependent on the exact radial range considered. Fig. \ref{fig:VPOSenhancement} plots the VPOS enhancement as a function of radius, split into the three angular momentum bins. The high initial angular momentum particles always show the least VPOS enhancement. In particular, they show no more than a 10\% enhancement within 200\,kpc from the host, the distance range covered by most MW satellites with reliably measured orbital poles. We also note that our overall VPOS enhancement is very comparable to that reported by \citetalias{GC21} for their fiducial model (see their Fig. 15), upon which we modelled our initial conditions. This confirms that our results are indeed comparable with their study.

It is to be expected that differences in initial conditions and model parameters, such as in the anisotropy parameter of the dark matter halo particles, the exact phase along the LMC-infall orbit at which the model is evaluated, or the mass ratio of the LMC and MW analog, all influence the degree of inferred orbital pole enhancement. For example, a likely source of differences ompared to \citetalias{GC21} is our choice of dark matter halo setup. While we deliberately chose one with isotropic orbits, the \citetalias{GC21} model implies somewhat more radially biased orbits. Since we find that particles on more radial orbits show stronger enhancements, a different mix of orbits will lead to differences in the overall degree of orbital pole enhancement. Furthermore, the simulations of GC21 and ours do not share the exact same orbits of the LMC analogs and are likely analyzed as slightly different phases of their orbits, which can be expected to propagate into further differences in the exact degree of orbital pole enhancement. 
Finally, we note that our simulation has a much lower particle resolution than that of \citetalias{GC21}. While the overall dynamics and effect of the LMC is certainly captured with our approach, it is possible that the lower resolution does not capture all the resonances of the dark matter halo. For example, while our simulation does show a wake induced by the LMC,  its detailed properties and behaviour might not yet be fully converged at our resolution. An increased resolution could thus potentially change the detailed results slightly.
It is reassuring how closely the overall degree of enhancement of the two studies agree despite these effects, with reported orbital pole enhancements in the range of 9 to 15\% in the two simulations.

\section{Conclusions} \label{sec:conclusions}

\citetalias{GC21} have demonstrated that the infall of a massive, LMC-like galaxy onto a MW-like host results in an enhancement of orbital poles of dark matter particles in the direction of the orbital plane of the interaction. For the MW, this direction aligns with the VPOS. This in turn led \citetalias{GC21} to propose that the infall of a massive LMC could have changed the orbital pole directions of the observed MW satellite galaxies, and that this could help explain the observed clustering or satellite orbital poles towards the VPOS normal. 

We confirm the overall results of \citetalias{GC21} with our own simulation of a LMC-MW like interaction. However, our results strongly disagree with their suggestion that this can help explain the origin of the VPOS. We find that accounting for the radius-dependent shifts in position and velocity of the MW halo center-of-mass induced by the LMC infall does not remove nor appreciably weaken the clustering of orbital poles towards the VPOS normal. This is likely due to the small relative COM shifts on the positions and velocities of the MW satellites. While the orbital pole direction of objects on more radial orbits (low angular momenta) can be substantially changed by even a small perturbation to their position or velocity, objects on less radial orbits (higher angular momenta) are much less affected by the same shifts. The observed MW satellites, as well as dark matter sub-halos in hydrodynamical cosmological simulations, have less radial orbits and higher tangential velocities (and thus higher specific angular momenta) than many dark matter halo particles or subhalos in dark-matter-only simulations \citep{2004MNRAS.352..535D, 2017MNRAS.468L..41C, 2017MNRAS.471.1709G, 2020ApJ...894...10L}. We hypothesize that the observed MW satellites have too high angular momenta to be sufficiently affected by the LMC infall to explain their strong clustering of orbital poles.

Our numerical simulation of an LMC-MW like interaction confirms this, while -- beyond the mere shift in positiona and the reflex motion -- also accounting for the additional effects of direct torques by the LMC and its induced wake. Within the distance range of 50 to 250\, kpc, which is most relevant for the observed MW satellites, the simulation predicts an orbital pole enhancement of 9\% in the VPOS direction. In the same area of sky, 12 out of 31 MW satellite orbital poles are found, corresponding to an enhancement of 300\% over the expected isotropic share of three orbital poles. Even excluding the two likely and one possible LMC satellites among these, which could artificially boost the enhancement because their orbits would be aligned with that of the LMC, results in an enhancement of 220\%. The observed signal is thus substantially larger than the one predicted to be induced by the LMC. This discrepancy increases further to almost two orders of magnitude (3\% vs. 220--300\%) if only particles with angular momenta as high as those of the MW satellites are considered. Particles that initially had very low angular momenta show the strongest enhancement in orbital pole density towards the VPOS normal. However, these angular momenta are inconsistent with the observed dwarf galaxies, as based on \textit{Gaia}-EDR3 proper motions \citep{2022A&A...657A..54B}. Restricting the analysis to particles of comparably high angular momenta, the orbital pole enhancement all but vanishes. The LMC-like infall is therefore unable to induce a pronounced enhancement of orbital poles in the VPOS direction for simulation particles with angular momenta comparable to those inferred for the observed MW dwarfs.

In summary, the LMC-induced overdensity of orbital poles is (i) only present for particles with low specific angular momenta, while the effect becomes fully negligible towards the higher specific angular momenta of the observed MW satellites, and (ii) is still too small an effect to account for the strong degree of orbital pole clustering among the observed MW satellites even if assuming the impact on them were as strong as on particles with low specific angular momenta. The LMC's infall onto the MW therefore does not suffice by a large margin to explain the observed VPOS.

If the LMC cannot explain the VPOS, its orbital alignment with the VPOS despite its very recent infall nevertheless remains a puzzling coincidence \citep{Pawlowski2013}. While alternative approaches have proposed some potential explanatory mechanisms \citep{1976MNRAS.174..695L, 2010ApJ...725L..24Y, 2018A&A...614A..59B}, at least within a classical $\Lambda$CDM framework this alignment appears to be nothing but a chance event.

\section*{Data Availability}

The initial conditions and final snapshot of the simulation described in Section~\ref{sec:Sim} can be obtained at the following address: \href{https://seafile.unistra.fr/d/f0ff2f519f4e42b792a0/}{https://seafile.unistra.fr/d/f0ff2f519f4e42b792a0/}.

\begin{acknowledgments}
MSP and ST acknowledge funding of a Leibniz-Junior Research Group (project number J94/2020) via the Leibniz Competition, MSP also thanks the Klaus Tschira Stiftung and German Scholars Organization for support via a KT Boost Fund. MSP and BF acknowledge support from the Partenariat Hubert Curien (PHC) for PROCOPE project 44677UE and the Deutscher Akademischer Austauschdienst for PPP grant 57512596 funded by the Bundesministerium fur Bildung und Forschung. PAO, BF and RI acknowledge funding from the European Research Council (ERC) under the European Unions Horizon 2020 research and innovation programme (grant agreement No. 834148) and from the Agence Nationale de la Recherche (ANR projects ANR-18-CE31-0006 and ANR-19-CE31-0017). 
This research has made use of NASA's Astrophysics Data System.
\end{acknowledgments}

\bibliography{LMCOrbitalPoles_bib}{}

\begin{thebibliography}{}
\expandafter\ifx\csname natexlab\endcsname\relax\def\natexlab#1{#1}\fi
\providecommand{\url}[1]{\href{#1}{#1}}
\providecommand{\dodoi}[1]{doi:~\href{http://doi.org/#1}{\nolinkurl{#1}}}
\providecommand{\doeprint}[1]{\href{http://ascl.net/#1}{\nolinkurl{http://ascl.net/#1}}}
\providecommand{\doarXiv}[1]{\href{https://arxiv.org/abs/#1}{\nolinkurl{https://arxiv.org/abs/#1}}}

\bibitem[{{Battaglia} {et~al.}(2022){Battaglia}, {Taibi}, {Thomas}, \&
  {Fritz}}]{2022A&A...657A..54B}
{Battaglia}, G., {Taibi}, S., {Thomas}, G.~F., \& {Fritz}, T.~K. 2022, \aap,
  657, A54, \dodoi{10.1051/0004-6361/202141528}

\bibitem[{{B{\'\i}lek} {et~al.}(2018){B{\'\i}lek}, {Thies}, {Kroupa}, \&
  {Famaey}}]{2018A&A...614A..59B}
{B{\'\i}lek}, M., {Thies}, I., {Kroupa}, P., \& {Famaey}, B. 2018, \aap, 614,
  A59, \dodoi{10.1051/0004-6361/201731939}

\bibitem[{{Binney} \& {Tremaine}(2008)}]{2008gady.book.....B}
{Binney}, J., \& {Tremaine}, S. 2008, {Galactic Dynamics: Second Edition}

\bibitem[{{Bullock} \& {Boylan-Kolchin}(2017)}]{2017ARA&A..55..343B}
{Bullock}, J.~S., \& {Boylan-Kolchin}, M. 2017, \araa, 55, 343,
  \dodoi{10.1146/annurev-astro-091916-055313}

\bibitem[{{Cautun} \& {Frenk}(2017)}]{2017MNRAS.468L..41C}
{Cautun}, M., \& {Frenk}, C.~S. 2017, \mnras, 468, L41,
  \dodoi{10.1093/mnrasl/slx025}

\bibitem[{{Correa Magnus} \& {Vasiliev}(2021)}]{2021arXiv211000018C}
{Correa Magnus}, L., \& {Vasiliev}, E. 2021, arXiv e-prints, arXiv:2110.00018.
\newblock \doarXiv{2110.00018}

\bibitem[{{Dehnen}(2000)}]{2000ApJ...536L..39D}
{Dehnen}, W. 2000, \apjl, 536, L39, \dodoi{10.1086/312724}

\bibitem[{{Diemand} {et~al.}(2004){Diemand}, {Moore}, \&
  {Stadel}}]{2004MNRAS.352..535D}
{Diemand}, J., {Moore}, B., \& {Stadel}, J. 2004, \mnras, 352, 535,
  \dodoi{10.1111/j.1365-2966.2004.07940.x}

\bibitem[{{Erkal} \& {Belokurov}(2020)}]{2020MNRAS.495.2554E}
{Erkal}, D., \& {Belokurov}, V.~A. 2020, \mnras, 495, 2554,
  \dodoi{10.1093/mnras/staa1238}

\bibitem[{{Fritz} {et~al.}(2018){Fritz}, {Battaglia}, {Pawlowski},
  {Kallivayalil}, {van der Marel}, {Sohn}, {Brook}, \&
  {Besla}}]{2018A&A...619A.103F}
{Fritz}, T.~K., {Battaglia}, G., {Pawlowski}, M.~S., {et~al.} 2018, \aap, 619,
  A103, \dodoi{10.1051/0004-6361/201833343}

\bibitem[{{Fritz} {et~al.}(2019){Fritz}, {Carrera}, {Battaglia}, \&
  {Taibi}}]{2019A&A...623A.129F}
{Fritz}, T.~K., {Carrera}, R., {Battaglia}, G., \& {Taibi}, S. 2019, \aap, 623,
  A129, \dodoi{10.1051/0004-6361/201833458}

\bibitem[{{Garavito-Camargo} {et~al.}(2021){Garavito-Camargo}, {Patel},
  {Besla}, {Price-Whelan}, {G{\'o}mez}, {Laporte}, \& {Johnston}}]{GC21}
{Garavito-Camargo}, N., {Patel}, E., {Besla}, G., {et~al.} 2021, \apj, 923,
  140, \dodoi{10.3847/1538-4357/ac2c05}

\bibitem[{{Garrison-Kimmel} {et~al.}(2017){Garrison-Kimmel}, {Wetzel},
  {Bullock}, {Hopkins}, {Boylan-Kolchin}, {Faucher-Gigu{\`e}re}, {Kere{\v{s}}},
  {Quataert}, {Sanderson}, {Graus}, \& {Kelley}}]{2017MNRAS.471.1709G}
{Garrison-Kimmel}, S., {Wetzel}, A., {Bullock}, J.~S., {et~al.} 2017, \mnras,
  471, 1709, \dodoi{10.1093/mnras/stx1710}

\bibitem[{{Hammer} {et~al.}(2021){Hammer}, {Wang}, {Pawlowski}, {Yang},
  {Bonifacio}, {Li}, {Babusiaux}, \& {Arenou}}]{2021arXiv210911557H}
{Hammer}, F., {Wang}, J., {Pawlowski}, M., {et~al.} 2021, arXiv e-prints,
  arXiv:2109.11557.
\newblock \doarXiv{2109.11557}

\bibitem[{{Heesters} {et~al.}(2021){Heesters}, {Habas}, {Marleau},
  {M{\"u}ller}, {Duc}, {Poulain}, {Durrell}, {S{\'a}nchez-Janssen}, \&
  {Paudel}}]{2021arXiv210810189H}
{Heesters}, N., {Habas}, R., {Marleau}, F.~R., {et~al.} 2021, arXiv e-prints,
  arXiv:2108.10189.
\newblock \doarXiv{2108.10189}

\bibitem[{{Ibata} {et~al.}(2013){Ibata}, {Lewis}, {Conn}, {Irwin},
  {McConnachie}, {Chapman}, {Collins}, {Fardal}, {Ferguson}, {Ibata}, {Mackey},
  {Martin}, {Navarro}, {Rich}, {Valls-Gabaud}, \&
  {Widrow}}]{2013Natur.493...62I}
{Ibata}, R.~A., {Lewis}, G.~F., {Conn}, A.~R., {et~al.} 2013, \nat, 493, 62,
  \dodoi{10.1038/nature11717}

\bibitem[{{Kallivayalil} {et~al.}(2013){Kallivayalil}, {van der Marel},
  {Besla}, {Anderson}, \& {Alcock}}]{2013ApJ...764..161K}
{Kallivayalil}, N., {van der Marel}, R.~P., {Besla}, G., {Anderson}, J., \&
  {Alcock}, C. 2013, \apj, 764, 161, \dodoi{10.1088/0004-637X/764/2/161}

\bibitem[{{Kelley} {et~al.}(2019){Kelley}, {Bullock}, {Garrison-Kimmel},
  {Boylan-Kolchin}, {Pawlowski}, \& {Graus}}]{2019MNRAS.487.4409K}
{Kelley}, T., {Bullock}, J.~S., {Garrison-Kimmel}, S., {et~al.} 2019, \mnras,
  487, 4409, \dodoi{10.1093/mnras/stz1553}

\bibitem[{{Kroupa} {et~al.}(2005){Kroupa}, {Theis}, \&
  {Boily}}]{2005A&A...431..517K}
{Kroupa}, P., {Theis}, C., \& {Boily}, C.~M. 2005, \aap, 431, 517,
  \dodoi{10.1051/0004-6361:20041122}

\bibitem[{{Kunkel} \& {Demers}(1976)}]{1976RGOB..182..241K}
{Kunkel}, W.~E., \& {Demers}, S. 1976, in The Galaxy and the Local Group, Vol.
  182, 241

\bibitem[{{Li} {et~al.}(2021){Li}, {Hammer}, {Babusiaux}, {Pawlowski}, {Yang},
  {Arenou}, {Du}, \& {Wang}}]{2021ApJ...916....8L}
{Li}, H., {Hammer}, F., {Babusiaux}, C., {et~al.} 2021, \apj, 916, 8,
  \dodoi{10.3847/1538-4357/ac0436}

\bibitem[{{Li} {et~al.}(2020){Li}, {Qian}, {Han}, {Li}, {Wang}, \&
  {Jing}}]{2020ApJ...894...10L}
{Li}, Z.-Z., {Qian}, Y.-Z., {Han}, J., {et~al.} 2020, \apj, 894, 10,
  \dodoi{10.3847/1538-4357/ab84f0}

\bibitem[{{Lynden-Bell}(1976)}]{1976MNRAS.174..695L}
{Lynden-Bell}, D. 1976, \mnras, 174, 695, \dodoi{10.1093/mnras/174.3.695}

\bibitem[{{Mart{\'\i}nez-Delgado} {et~al.}(2021){Mart{\'\i}nez-Delgado},
  {Makarov}, {Javanmardi}, {Pawlowski}, {Makarova}, {Donatiello}, {Lang},
  {Rom{\'a}n}, {Vivas}, \& {Carballo-Bello}}]{2021A&A...652A..48M}
{Mart{\'\i}nez-Delgado}, D., {Makarov}, D., {Javanmardi}, B., {et~al.} 2021,
  \aap, 652, A48, \dodoi{10.1051/0004-6361/202141242}

\bibitem[{{M{\"u}ller} {et~al.}(2018){M{\"u}ller}, {Pawlowski}, {Jerjen}, \&
  {Lelli}}]{2018Sci...359..534M}
{M{\"u}ller}, O., {Pawlowski}, M.~S., {Jerjen}, H., \& {Lelli}, F. 2018,
  Science, 359, 534, \dodoi{10.1126/science.aao1858}

\bibitem[{{M{\"u}ller} {et~al.}(2021){M{\"u}ller}, {Pawlowski}, {Lelli},
  {Fahrion}, {Rejkuba}, {Hilker}, {Kanehisa}, {Libeskind}, \&
  {Jerjen}}]{2021A&A...645L...5M}
{M{\"u}ller}, O., {Pawlowski}, M.~S., {Lelli}, F., {et~al.} 2021, \aap, 645,
  L5, \dodoi{10.1051/0004-6361/202039973}

\bibitem[{{Patel} {et~al.}(2020){Patel}, {Kallivayalil}, {Garavito-Camargo},
  {Besla}, {Weisz}, {van der Marel}, {Boylan-Kolchin}, {Pawlowski}, \&
  {G{\'o}mez}}]{2020ApJ...893..121P}
{Patel}, E., {Kallivayalil}, N., {Garavito-Camargo}, N., {et~al.} 2020, \apj,
  893, 121, \dodoi{10.3847/1538-4357/ab7b75}

\bibitem[{{Paudel} {et~al.}(2021){Paudel}, {Yoon}, \&
  {Smith}}]{2021ApJ...917L..18P}
{Paudel}, S., {Yoon}, S.-J., \& {Smith}, R. 2021, \apjl, 917, L18,
  \dodoi{10.3847/2041-8213/ac1866}

\bibitem[{{Pawlowski}(2018)}]{2018MPLA...3330004P}
{Pawlowski}, M.~S. 2018, Modern Physics Letters A, 33, 1830004,
  \dodoi{10.1142/S0217732318300045}

\bibitem[{{Pawlowski} {et~al.}(2013){Pawlowski}, {Kroupa}, \&
  {Jerjen}}]{Pawlowski2013}
{Pawlowski}, M.~S., {Kroupa}, P., \& {Jerjen}, H. 2013, \mnras, 435, 1928,
  \dodoi{10.1093/mnras/stt1384}

\bibitem[{{Pawlowski} {et~al.}(2012){Pawlowski}, {Pflamm-Altenburg}, \&
  {Kroupa}}]{2012MNRAS.423.1109P}
{Pawlowski}, M.~S., {Pflamm-Altenburg}, J., \& {Kroupa}, P. 2012, \mnras, 423,
  1109, \dodoi{10.1111/j.1365-2966.2012.20937.x}

\bibitem[{{Pe{\~n}arrubia} {et~al.}(2016){Pe{\~n}arrubia}, {G{\'o}mez},
  {Besla}, {Erkal}, \& {Ma}}]{2016MNRAS.456L..54P}
{Pe{\~n}arrubia}, J., {G{\'o}mez}, F.~A., {Besla}, G., {Erkal}, D., \& {Ma},
  Y.-Z. 2016, \mnras, 456, L54, \dodoi{10.1093/mnrasl/slv160}

\bibitem[{{Petersen} \& {Pe{\~n}arrubia}(2021)}]{2021NatAs...5..251P}
{Petersen}, M.~S., \& {Pe{\~n}arrubia}, J. 2021, Nature Astronomy, 5, 251,
  \dodoi{10.1038/s41550-020-01254-3}

\bibitem[{{Riley} {et~al.}(2019){Riley}, {Fattahi}, {Pace}, {Strigari},
  {Frenk}, {G{\'o}mez}, {Grand}, {Marinacci}, {Navarro}, {Pakmor}, {Simpson},
  \& {White}}]{2019MNRAS.486.2679R}
{Riley}, A.~H., {Fattahi}, A., {Pace}, A.~B., {et~al.} 2019, \mnras, 486, 2679,
  \dodoi{10.1093/mnras/stz973}

\bibitem[{{Rozier} {et~al.}(2022){Rozier}, {Famaey}, {Siebert}, {Monari},
  {Pichon}, \& {Ibata}}]{2022arXiv220105589R}
{Rozier}, S., {Famaey}, B., {Siebert}, A., {et~al.} 2022, arXiv e-prints,
  arXiv:2201.05589.
\newblock \doarXiv{2201.05589}

\bibitem[{{Teuben}(1995)}]{1995ASPC...77..398T}
{Teuben}, P. 1995, in Astronomical Society of the Pacific Conference Series,
  Vol.~77, Astronomical Data Analysis Software and Systems IV, ed. R.~A.
  {Shaw}, H.~E. {Payne}, \& J.~J.~E. {Hayes}, 398

\bibitem[{{Vasiliev}(2019)}]{2019MNRAS.482.1525V}
{Vasiliev}, E. 2019, \mnras, 482, 1525, \dodoi{10.1093/mnras/sty2672}

\bibitem[{{Vasiliev} {et~al.}(2021){Vasiliev}, {Belokurov}, \&
  {Erkal}}]{2021MNRAS.501.2279V}
{Vasiliev}, E., {Belokurov}, V., \& {Erkal}, D. 2021, \mnras, 501, 2279,
  \dodoi{10.1093/mnras/staa3673}

\bibitem[{{Yang} \& {Hammer}(2010)}]{2010ApJ...725L..24Y}
{Yang}, Y., \& {Hammer}, F. 2010, \apjl, 725, L24,
  \dodoi{10.1088/2041-8205/725/1/L24}

\end{thebibliography}
\bibliographystyle{aasjournal}



\end{document}